\documentclass[onefignum,onetabnum]{siamonline190516}

\usepackage{tikz}\usetikzlibrary{calc,patterns,
                 decorations.pathmorphing,
                 decorations.markings,
                 circuits.ee.IEC}
\usepackage{graphicx,epstopdf}
\usepackage[caption=false]{subfig}
\usepackage{amssymb}
\usepackage{algorithmic} 
\Crefname{ALC@unique}{Line}{Lines} 
\usepackage{bold-extra}

\begin{document}

\title{Frequency Response and Transfer Functions of Large Self-similar Networks\thanks{Submitted to the editors DATE.
\funding{The support of the US National Science Foundation under Grant No. CMMI 1826079 is gratefully acknowledged.}}}
\author{Xiangyu Ni\thanks{email: xni@nd.edu}
\and Bill Goodwine\thanks{email: billgoodwine@nd.edu}}
\headers{Frequency Response and Transfer Functions of Large Self-similar Networks}{Xiangyu Ni, and Bill Goodwine}

\maketitle

\begin{abstract}
This paper focuses on computing the frequency response and transfer functions for large self-similar networks under different circumstances. Modeling large scale systems is difficult due, typically, to the dimension of the problem, and self-similarity is the characteristic we exploit to make the problem more tractable.  For each circumstance, we propose algorithms to obtain both transfer functions and frequency response, and we show that finite networks' dynamics are integer order, while infinite networks are fractional order or irrational. Based on that result, we also show that the effect of varying a network's operating condition to its dynamics can always be isolated, which is then expressed as a multiplicative disturbance acting upon a nominal plant. In addition, we analyze the non-integer-order nature residing in infinite dimensional systems in the context of self-similar networks. Finally, leveraging the main result of this paper, we also illustrate its capability of approximating some irrational expressions by using rational functions.
\end{abstract}



\section{Introduction}
Large self-similar networks and fractal geometries appear in many applications. Some are natural, such as river networks whose configuration is studied in \cite{5597994}. Other natural networks exist in biology. For instance, the methods of determining the fractal dimension for the branching vascular tree in human retina and human bronchial tree are reviewed in \cite{Masters2004Fractal}. Infinite mechanical ladder networks and fractal networks are used in \cite{IONESCU2017433} as two of the modeling tools for the mechanical properties in lungs. Some artificial networks can be seen in implementations such as robotic swarms, power grids, and ventilating systems. For example, some heat exchangers designed with fractal geometries are reviewed in \cite{HUANG20171}. The utility of the fractal tree-shaped fin in the energy charging performance is explored in \cite{doi:10.1002/er.5268}.

Because of the many and varied useful applications, there has been significant research attention focused on these types of large scale networks. For natural networks, a thermal conductivity model of biological tissue which includes the effects of the vascular network's geometry and the blood flow's convection is proposed in \cite{LI2014219}. The effects on flow behavior brought by changing the geometry and the operating condition of biologically inspired networks are studied in \cite{10.1115/IMECE2017-71681}. For robotic swarms, an approach to adapt their aggregation behavior to the variations in the swarm density and the external environment is present in \cite{doi:10.1162/isal}. For power grids, a nonlinear model for preventing cascading failures in complex networks is suggested in \cite{Chen2020}. A fast computation method for the electrical characteristics such as currents, voltages, impedance of a complex electrical ladder network is developed in \cite{doi:10.1002/cta.2446}. The capacitance between any two vertices inside infinite triangular and honeycomb networks is studied in \cite{Owaidat2014The}. For thermal networks, a compact model which can quickly simulate their physical behavior in various operating conditions is discussed in \cite{GUELPA2019998}.

In this paper, we focus on computing large self-similar networks' frequency response and transfer functions which quantitatively predict a dynamical system's output in response to a variety of stimuli. Our main motivation is that large networks can then be further analyzed or controlled by available frequency-domain tools given the results obtained by the methods proposed in this work. Existing literature regarding evaluating complicated systems' dynamics is often data driven. The fractional-order transfer function of a voltammetric electronic tongue system is approximated through identification in \cite{KUMAR2020107064}. The impedance response of NaCl-glucose solutions is identified from measurements in \cite{OLARTE2014213}. 

In contrast, the approach in this paper is model based and aims at the exact computation of a large network's dynamics under different conditions. The gap filled by this work can be observed from \Cref{tab:literature}, which classifies a large network by two ways. First, a large network is either undamaged or damaged, depending on all of its components' statuses. If all components' constants are same as their \textit{undamaged constants}, that network is called undamaged. Otherwise, it is damaged.\footnote{One of the main motivations for this work is health monitoring and the damage detection problem.} Second, a network is either finite or infinite relying on its number of generations. \Cref{tab:literature} shows that the main contribution of this paper is twofold. First, the algorithms for frequency response in this paper are modular and thus are designed for any self-similar networks which satisfy the assumptions in \Cref{sec:assum}, while existing literature often concentrates on some specific networks individually. Second, this paper also proposes algorithms to compute transfer functions for large networks in more general cases. In contrast, existing literature only derives those for infinite undamaged networks \cite{goodwine2014modeling,heymans1994fractal,mayes2012reduction}. In this paper, we use the term \emph{transfer function} to indicate the expression of the ratio between a network's output signal and its input signal in the frequency domain, which is denoted by $G(s)$. On the other hand, the term \emph{frequency response} is for the numerical value of the corresponding $G(i\omega)$ at some angular frequencies $\omega$.
\begin{table}
    \footnotesize
    \caption{Existing literature regarding exact computing large networks' dynamics.}\label{tab:literature}
    \begin{center}
    \begin{tabular}{|c|c|c|} \hline
     & Infinite & Finite \\ \hline
    Undamaged & Transfer functions \& Frequency response & Frequency response \\
    Damaged & Frequency response & Frequency response \\ \hline
    \end{tabular}
    \end{center}
\end{table}

Knowing large networks' transfer functions gives this paper three additional contributions. First, we show that the effect of varying a network's operating condition on its transfer function can be isolated from the one before that condition change. The isolation is presented as a multiplicative disturbance in this paper, which is a classical model in the robust control area. Second, we observe that finite networks' transfer functions are always integer-order, while those of infinite networks are fractional or even irrational. Therefore, that difference offers one concrete example confirming the fact that non-integer-order dynamics is brought naturally by infinite dimensional systems \cite{YamamotoNov1988}. Third, because the dynamics of a finite network converges to that of an infinite network, we can thus use those rational transfer functions of finite networks to approximate the irrational expressions inside infinite networks' transfer functions, which is similar to a Pad{\'e} approximation. Other approximation methods in literature are compared in \cite{5698677}. Another similar example can be found in \cite{Kasperski_2013} which shows how to realize any irrational value of resistance by a finite electrical network with unit resistors.

The rest of this paper is structured as follows. \Cref{sec:assum} lists assumptions which a network needs to fulfill for the methods presented in this paper to apply. In addition, it also introduces three examples which we are going to showcase throughout this paper. \Cref{sec:finMod} and \Cref{sec:infMod} propose algorithms to compute frequency response and transfer functions of finite and infinite networks, respectively. \Cref{sec:dis} discusses applications brought by the knowledge of networks' dynamics obtained by the methods from this paper. Finally, \Cref{sec:con} concludes this paper.

\section{Assumptions and example networks}
\label{sec:assum}
As mentioned in the Introduction, the goal of this paper is to compute both frequency response and transfer function of a large self-similar network. That refers to a ratio of two frequency-domain quantities at the same point within a network, such as the input impedance of an electrical network. For a network to be qualified for the approach proposed in this paper, it must satisfy the following assumptions.
\begin{itemize}
    \item (A-1) The network is one-dimensional.
    \item (A-2) The network is self-similar \cite{mandelbrot1983fractal}. That is, the structure of the network repeats after every certain number of generations.
    \item (A-3) All components within the network are connected either in parallel or in series.
    \item (A-4) All components are linear in the frequency domain with time-invariant proportional constants, such as idealized dampers, idealized capacitors, etc.
\end{itemize}

\begin{figure}
    \centering
    \begin{tikzpicture}
        \tikzstyle{spring}=[thick,decorate,decoration={zigzag,pre length=0.3cm,post length=0.3cm,segment length=6}]
        \tikzstyle{damper}=[thick,decoration={markings,mark connection node=dmp,mark=at position 0.5 with 
        {
            \node (dmp) [thick,inner sep=0pt,transform shape,rotate=-90,minimum width=15pt,minimum height=3pt,draw=none] {};
            \draw [thick] ($(dmp.north east)+(2pt,0)$) -- (dmp.south east) -- (dmp.south west) -- ($(dmp.north west)+(2pt,0)$);
            \draw [thick] ($(dmp.north)+(0,-5pt)$) -- ($(dmp.north)+(0,5pt)$);
        }
        },decorate]
        \node at (0,0) (leftF) {$f$};
        \filldraw[black] (0.7,0) circle (2pt) node [above] {$x_{1,1}$};
        \draw[thick, -latex] (leftF.east) -- (0.7,0);
        \draw[thick] (0.7,0) -- (1.1,0);
        \draw[thick] (1.1,1) -- (1.1,-1);
        \draw[spring] (1.1,1) -- (2.6,1) node [midway,above=1pt] {$k_{1,1}$};
        \draw[damper] (1.1,-1) -- (2.6,-1) node [midway,above=6pt] {$b_{1,1}$};
        \draw (2.6,1) circle (2pt) node [above] {$x_{2,1}$};
        \draw (2.6,-1) circle (2pt) node [above] {$x_{2,2}$};
        \draw[thick] (2.6,1) -- (3,1);
        \draw[thick] (2.6,-1) -- (3,-1);
        \draw[thick] (3,1.5) -- (3,0.5);
        \draw[thick] (3,-0.5) -- (3,-1.5);
        \draw[spring] (3,1.5) -- (4.5,1.5) node [midway,above=1pt] {$k_{2,1}$};
        \draw[damper] (3,0.5) -- (4.5,0.5) node [midway,above=6pt] {$b_{2,1}$};
        \draw[spring] (3,-0.5) -- (4.5,-0.5) node [midway,above=1pt] {$k_{2,2}$};
        \draw[damper] (3,-1.5) -- (4.5,-1.5) node [midway,above=6pt] {$b_{2,2}$};
        \draw (4.5,1.5) circle (2pt) node [above] {$x_{3,1}$};
        \draw (4.5,0.5) circle (2pt) node [above] {$x_{3,2}$};
        \draw (4.5,-0.5) circle (2pt) node [above] {$x_{3,3}$};
        \draw (4.5,-1.5) circle (2pt) node [above] {$x_{3,4}$};
        \draw[thick] (4.5,1.5) -- (5,1.5);
        \draw[thick] (4.5,0.5) -- (5,0.5);
        \draw[thick] (4.5,-0.5) -- (5,-0.5);
        \draw[thick] (4.5,-1.5) -- (5,-1.5);
        \draw[thick] (5,2) -- (5,1);
        \draw[spring] (5,2) -- (6.5,2) node [midway,above=1pt] {$k_{3,1}$};
        \draw[damper] (5,1) -- (6.5,1) node [midway,above=6pt] {$b_{3,1}$};
        \draw (6.5,2) circle (2pt) node [above] {$x_{4,1}$};
        \draw (6.5,1) circle (2pt) node [above] {$x_{4,2}$};
        \node at (7,2) {$\cdots$};
        \node at (7,1) {$\cdots$};
        \node at (5.5,0.5) {$\cdots$};
        \node at (5.5,-0.5) {$\cdots$};
        \node at (5.5,-1.5) {$\cdots$};
        \draw[black, very thick] (7.3,2.5) rectangle (7.5,-2) node [midway,below=2.4cm] {$x_{\text{last}}$};
        \filldraw[black] (7.4,2.4) circle (2pt);
        \filldraw[black] (7.4,2.2) circle (2pt);
        \filldraw[black] (7.4,2) circle (2pt);
        \filldraw[black] (7.4,1.8) circle (2pt);
        \filldraw[black] (7.4,1.6) circle (2pt);
        \filldraw[black] (7.4,1.4) circle (2pt);
        \node at (7.4,1.2) {$\vdots$};
        \node at (7.4,0.7) {$\vdots$};
        \node at (8,0) (rightF) {$f$};
        \draw[thick, -latex] (rightF.west) -- (7.5,0);
    \end{tikzpicture}
    \caption{Mechanical \textit{tree} network without intermediate masses}
    \label{fig:tree}
\end{figure}
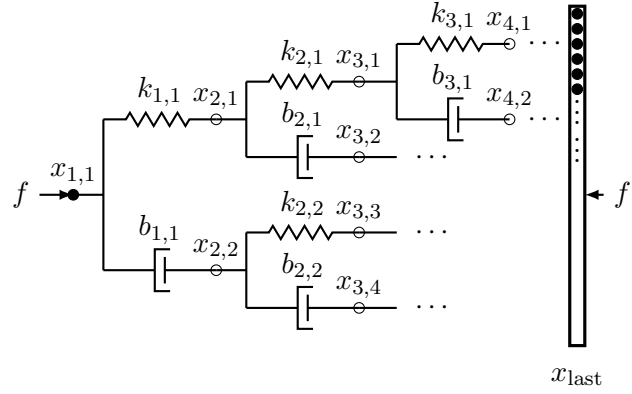

There are three example networks satisfying the above assumptions which we will refer to throughout this paper. The first example is a mechanical \textit{tree} network, as shown in \Cref{fig:tree}, consisting of linear springs and dampers but without masses at intermediate nodes, which has been used to model the relaxation of the aortic valve \cite{todd2005fractional} and viscoelastic behaviors \cite{heymans1994fractal} in literature. When undamaged, all spring constants $k_{1,1}=k_{2,1}=\cdots=k$ and all damper constants $b_{1,1}=b_{2,1}=\cdots=b$. Otherwise, the tree is damaged. The dynamics of interest in this paper for the tree is the ratio of its length, $X_{1,1}(s)-X_{\text{last}}(s)$, to the force exerted at both ends, $F(s)$. In our previous work \cite{ni2020damage}, we used the mathematical induction to prove that, regardless of whether it is damaged, the transfer function of an infinite tree always has the following formulation,
\begin{equation}
    G_\infty(s)=\frac{X_{1,1}(s)-X_{\text{last}}(s)}{F(s)}=\frac{1}{\sqrt{kbs}}\frac{s^n+c_{N,1}s^{n-\frac{1}{2}}+\cdots+c_{N,2n}}{s^n+c_{D,1}s^{n-\frac{1}{2}}+\cdots+c_{D,2n}}.
    \label{eq:infTreeTF}
\end{equation}
Using mathematical induction to achieve that result renders it difficult to be generalized to other large networks. Therefore, that previous work \cite{ni2020damage} only focused on the tree network. In contrast, this paper takes a more systematic route, so it can easily be extended to a general class of large networks that satisfy the assumptions (A-1) to (A-4). That capability of extension is shown by applying the approach proposed in this paper to the other two following examples. Note that the transfer function for the infinite tree~\cref{eq:infTreeTF} is fractional, which can be observed from the half orders of $s$. As the rest of this paper shows, that non-integer-order dynamics is a common feature for infinite networks.

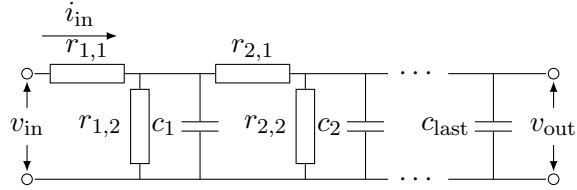
\begin{figure}
    \centering
    \begin{tikzpicture}[circuit ee IEC]
        \draw (0,0) circle (2pt);
        \draw (0.1,0) to [resistor] node [midway,above=1pt] {$r_{1,1}$} (1.5,0);
        \draw (1.5,0) to [resistor] node [midway,left=3pt] {$r_{1,2}$} (1.5,-1.4);
        \draw (2.3,0) to [capacitor] node [midway,left=5pt] {$c_1$} (2.3,-1.4);
        \draw[thin] (1.5,0) -- (2.3,0);
        \draw (2.3,0) to [resistor] node [midway,above=1pt] {$r_{2,1}$} (3.7,0);
        \draw (3.7,0) to [resistor] node [midway,left=3pt] {$r_{2,2}$} (3.7,-1.4);
        \draw (4.5,0) to [capacitor] node [midway,left=5pt] {$c_2$} (4.5,-1.4);
        \draw[thin] (3.7,0) -- (4.8,0);
        \draw (0,-1.4) circle (2pt);
        \draw[thin] (0.1,-1.4) -- (4.8,-1.4);
        \node[outer sep=0pt,thin] at (5.2,0) {$\cdots$};
        \node[outer sep=0pt,thin] at (5.2,-1.4) {$\cdots$};
        \draw (6.2,0) to [capacitor] node [midway,left=5pt] {$c_{\text{last}}$} (6.2,-1.4);
        \draw[thin] (5.6,0) -- (6.9,0);
        \draw[thin] (5.6,-1.4) -- (6.9,-1.4);
        \draw (7,0) circle (2pt);
        \draw (7,-1.4) circle (2pt);
        \node at (0,-0.7) (vi) {$v_{\text{in}}$};
        \node at (7,-0.7) (vo) {$v_{\text{out}}$};
        \draw[thin, -latex] (vi.north) -- (0,-0.1);
        \draw[thin, -latex] (vi.south) -- (0,-1.3);
        \draw[thin, -latex] (vo.north) -- (7,-0.1);
        \draw[thin, -latex] (vo.south) -- (7,-1.3);
        \draw[thin, -latex] (0.2,0.5) -- node [midway,above=0.1] {$i_{\text{in}}$} (1.2,0.5);
    \end{tikzpicture}
    \caption{\textit{Electrical ladder} network}
    \label{fig:eLadder}
\end{figure}

\Cref{fig:eLadder} shows the second example, the \textit{electrical ladder} network containing linear resistors and linear capacitors, which is a simplified version inspired by the model used in \cite{zhao2009simulation} for the ZPW-2000A high-speed railway track circuit system. Similar to the tree, when the electrical ladder is undamaged, all resistors $r_{1,1}=r_{2,1}=\cdots=r_1$, $r_{1,2}=r_{2,2}=\cdots=r_2$ and all capacitors $c_1=c_2=\cdots=c$. The frequency response and transfer function of interest in this paper for the electrical ladder network are its input impedance, $V_{\text{in}}(s)/I_{\text{in}}(s)$. One interpretation of the literature cited above is that when a train occupies a section of track, its presence can be detected in a manner similar to damage in the network, \textit{i.e.}, some properties of the compontents are changed. 

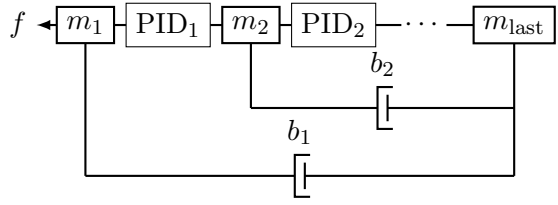
\begin{figure}
    \centering
    \begin{tikzpicture}
    \tikzstyle{spring}=[thick,decorate,decoration={zigzag,pre length=0.3cm,post length=0.3cm,segment length=6}]
        \tikzstyle{damper}=[thick,decoration={markings,mark connection node=dmp,mark=at position 0.5 with 
        {
            \node (dmp) [thick,inner sep=0pt,transform shape,rotate=-90,minimum width=15pt,minimum height=3pt,draw=none] {};
            \draw [thick] ($(dmp.north east)+(2pt,0)$) -- (dmp.south east) -- (dmp.south west) -- ($(dmp.north west)+(2pt,0)$);
            \draw [thick] ($(dmp.north)+(0,-5pt)$) -- ($(dmp.north)+(0,5pt)$);
        }
        },decorate]
        \node at (0,0) (f) {$f$};
        \node[draw,outer sep=0pt,thick] (m1) at (0.9,0) {$m_1$};
        \node[draw,outer sep=0pt,thin] (c1) at (2,0) {$\text{PID}_1$};
        \node[draw,outer sep=0pt,thick] (m2) at (3.1,0) {$m_2$};
        \node[draw,outer sep=0pt,thin] (c2) at (4.2,0) {$\text{PID}_2$};
        \node[outer sep=0pt,thick] at (5.4,0) {$\cdots$};
        \node[draw,outer sep=0pt,thick] (mL) at (6.6,0) {$m_{\text{last}}$};
        \draw[damper] (0.9,-2) -- (6.6,-2) node [midway,above=8pt] {$b_1$};
        \draw[damper] (3.1,-1.1) -- (6.6,-1.1) node [midway,above=8pt] {$b_2$};
        \draw[thick] (m1.south) -- (0.9,-2);
        \draw[thick] (m2.south) -- (3.1,-1.1);
        \draw[thick] (mL.south) -- (6.6,-2);
        \draw[thick, -latex] (m1.west) -- (f.east);
        \draw[thick] (m1.east) -- (c1.west);
        \draw[thick] (c1.east) -- (m2.west);
        \draw[thick] (m2.east) -- (c2.west);
        \draw[thick] (c2.east) -- (5.1,0);
        \draw[thick] (5.7,0) -- (mL.west);
    \end{tikzpicture}
    \caption{\textit{Mechanical ladder} network with masses}
    \label{fig:mLadder}
\end{figure}

The last example is the \textit{mechanical ladder} network with masses, as shown in \Cref{fig:mLadder}. This network can be viewed as a line of vehicles moving together in the same direction where each car knows the status of its neighboring cars and their distances are maintained by PID controllers. Every controller $\text{PID}_j$ exerts an equal amount of force $f_{\text{PID}_j}$ on the masses $m_j$ and $m_{j+1}$ with the opposite directions where
\begin{equation*}
    F_{\text{PID}_j}(s)=\left( k_{pj}+\frac{k_{ij}}{s}+k_{dj}s \right)X_j(s),
\end{equation*}
and $X_j(s)$ is the distance between the masses $m_j$ and $m_{j+1}$ in the frequency domain. Additionally, each car also knows the speed of the end car $m_{\text{last}}$ and attempts to follow that speed through a damper-like controller $b_j$. Note that in real applications, we may be more interested about the case where the leading vehicle is followed by the rest. However, in this paper, we swap that direction so that it is consistent with the other two examples. The force $f$ acts as a disturbance at the first car $m_1$. All masses are the same with $m=1kg$, and similar to the other examples, when undamaged, all PID constants $k_{pj}$, $k_{ij}$, and $k_{dj}$ are the same as their undamaged constants $k_p$, $k_i$, $k_d$ and all the damper constants $b_1=b_2=\cdots=b$. The dynamics of interest here is the ratio $X(s)/F(s)$ where $X(s)$ indicates the length of the entire mechanical ladder network between $m_1$ and $m_{\text{last}}$. For this mechanical ladder network, we have a specific assumption that the last vehicle $m_{\text{last}}$ is moving at a constant speed all the time. This can be achieved by using an external mechanism controlling the $m_{\text{last}}$, and/or by the fact that the impact of $f$ on $m_{\text{last}}$ is negligible, \textit{e.g.} $f$ is a gentle disturbance and the network's size is substantial.

We assume that components' damage can be represented mathematically in a multiplicative manner, and we use a pair of two lists, $(\boldsymbol{l},\boldsymbol{\epsilon})$, to denote a specific \textit{damage case} where $\boldsymbol{l}$ is the list of damaged components and $\boldsymbol{\epsilon}$ is the corresponding list of damage amounts. Taking a network consisting of purely springs as an example, if the undamaged spring constant is $k$, the damage case
\begin{equation*}
    (\boldsymbol{l},\boldsymbol{\epsilon})=([k_1,k_2],[0.1,0.2])
\end{equation*}
means that the components $k_1$ and $k_2$ are damaged, and their constants become $k_1=0.1k$ and $k_2=0.2k$, while all the other springs are unchanged with their constants staying at $k$. Note that when an element in $\boldsymbol{\epsilon}$ is close to 0, its corresponding component undergoes severe damage. In contrast, if it is close to 1, that component's damage is slight. Furthermore, we use $G_{g,(\boldsymbol{l},\boldsymbol{\epsilon})}(s)$ to distinguish the transfer functions for the different cases. The positive integer $g$ indicates the number of generations in the network. When $g$ is $\infty$, the network has an infinite number of generations. When the damage case $(\boldsymbol{l},\boldsymbol{\epsilon})$ is $\varnothing$, the formulation is for an undamaged network. Especially, when there are no contents at this entry, the formulation is a general one which works for both damaged and undamaged networks.

\section{Finite networks}
\label{sec:finMod}

In this section, we show how to exactly compute the frequency response and transfer functions for finite self-similar networks. The key ingredient is the recurrence formula which utilizes the self-similarity and relates the entire network to its sub-networks, whose derivation is in \Cref{sec:finModRec}. For finite networks, all computations repeatedly use the recurrence formula to gradually build up the entire network's frequency response. The challenge is then to plug the components' constants into that procedure at their correct iterations, which is automatically taken care by our proposed recursive algorithm in \Cref{sec:finModNum}. Building upon that, in \Cref{sec:finModAna}, we reveal how to convert the recurrence formula using convolutions to obtain a finite network's transfer function.

\subsection{Recurrence formula}
\label{sec:finModRec}
The basic idea of obtaining the recurrence formula for a network is deriving its transfer function in terms of its sub-networks' transfer functions. We use $G_{si}(s)$ to denote the transfer function of the sub-network $i$ and we use $G_r(s)$ to indicate a network's recurrence formula. 
The construction of the recurrence formula is shown below for each of those three examples.

\begin{figure}
    \centering
    \begin{tikzpicture}
        \tikzstyle{spring}=[thick,decorate,decoration={zigzag,pre length=0.3cm,post length=0.3cm,segment length=6}]
        \tikzstyle{damper}=[thick,decoration={markings,mark connection node=dmp,mark=at position 0.5 with 
        {
            \node (dmp) [thick,inner sep=0pt,transform shape,rotate=-90,minimum width=15pt,minimum height=3pt,draw=none] {};
            \draw [thick] ($(dmp.north east)+(2pt,0)$) -- (dmp.south east) -- (dmp.south west) -- ($(dmp.north west)+(2pt,0)$);
            \draw [thick] ($(dmp.north)+(0,-5pt)$) -- ($(dmp.north)+(0,5pt)$);
        }
        },decorate]
        \node at (0,0) (leftF) {$f$};
        \filldraw[black] (0.7,0) circle (2pt) node [above] {$x_{1,1}$};
        \draw[thick, -latex] (leftF.east) -- (0.7,0);
        \draw[thick] (0.7,0) -- (1.1,0);
        \draw[thick] (1.1,0.5) -- (1.1,-0.5);
        \draw[spring] (1.1,0.5) -- (2.5,0.5) node [midway,above=1pt] {$k_{1,1}$};
        \draw[damper] (1.1,-0.5) -- (2.5,-0.5) node [midway,above=6pt] {$b_{1,1}$};
        \draw (2.5,0.5) circle (2pt) node [above] {$x_{2,1}$};
        \draw (2.5,-0.5) circle (2pt) node [above] {$x_{2,2}$};
        \node[draw,outer sep=0pt,thick] (M1) at (3.5,0.5) {$G_{s1}(s)$};
        \node[draw,outer sep=0pt,thick] (M2) at (3.5,-0.5) {$G_{s2}(s)$};
        \draw[thick] (2.5,0.5) -- (M1.west);
        \draw[thick] (2.5,-0.5) -- (M2.west);
        \draw[black, thick] (4.4,1.2) rectangle (4.5,-1.2) node [midway,below=1.2cm] {$x_{\text{last}}$};
        \draw[thick] (M1.east) -- (4.4,0.5);
        \draw[thick] (M2.east) -- (4.4,-0.5);
        \node at (5.2,0) (rightF) {$f$};
        \draw[thick, -latex] (rightF.west) -- (4.5,0);
    \end{tikzpicture}
    \caption{An illustration for obtaining the tree network's recurrence formula}
    \label{fig:recTree}
\end{figure}
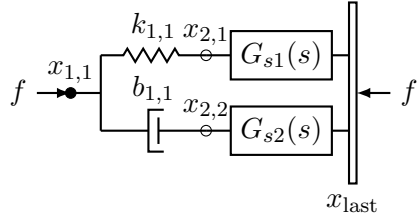

For the tree network in \Cref{fig:tree}, we assume its two sub-trees' transfer functions are known. That is, we know
\begin{align*}
    G_{s1}(s)&=\frac{X_{2,1}(s)-X_{\text{last}}(s)}{F_1(s)};\\
    G_{s2}(s)&=\frac{X_{2,2}(s)-X_{\text{last}}(s)}{F_2(s)};\\
    F(s)&=F_1(s)+F_2(s).
\end{align*}
Our goal is to represent the entire tree's transfer function
\begin{equation*}
G_r(s)=\frac{X_{1,1}(s)-X_{\text{last}}(s)}{F(s)},
\end{equation*}
in terms of $G_{s1}(s)$ and $G_{s2}(s)$. Using the series and parallel connection rules for idealized mechanical components, from the illustration in \Cref{fig:recTree}, it is straightforward to see that the tree network's recurrence formula is
\begin{align}
    &G_r(s)=\cfrac{1}{\cfrac{1}{\cfrac{1}{k_{1,1}}+G_{s1}(s)}+\cfrac{1}{\cfrac{1}{b_{1,1}s}+G_{s2}(s)}}\nonumber\\
    &=\frac{k_{1,1}b_{1,1}sG_{s1}(s)G_{s2}(s)+k_{1,1}G_{s1}(s)+b_{1,1}sG_{s2}(s)+1}{k_{1,1}b_{1,1}s(G_{s1}(s)+G_{s2}(s))+k_{1,1}+b_{1,1}s}.\label{eq:recTree}
\end{align}
The modeling procedure of the entire tree would employ the above recurrence formula $G_r(s)$, starting with the transfer function where the tree only has one generation,
\begin{equation}
    G_1(s)=\frac{1}{k_{1,1}+b_{1,1}s}.\label{eq:recTree1}
\end{equation}

For the electrical ladder network in \Cref{fig:eLadder}, its sketch for obtaining the recurrence formula is shown in \Cref{fig:recELadder}. In this instance, we assume to know the impedance for the sub-network which is
\begin{equation*}
    G_{s1}(s)=\frac{V_1(s)}{I_1(s)}.
\end{equation*}
We would like to derive the input impedance for the entire network
\begin{equation*}
    G_r(s)=\frac{V_{\text{in}}(s)}{I_{\text{in}}(s)}
\end{equation*}
in terms of $G_{s1}(s)$. Similarly, by using the series and parallel connection rules for idealized electrical components, we see that the recurrence formula for the electrical ladder network is
\begin{equation}
    G_r(s)=r_{1,1}+\cfrac{1}{\cfrac{1}{r_{1,2}}+c_1s+\cfrac{1}{G_{s1}(s)}}=\frac{ \left( r_{1,1}r_{1,2}c_1s+r_{1,1}+r_{1,2} \right) G_{s1}(s)+r_{1,1}r_{1,2}}{(r_{1,2}c_1s+1)G_{s1}(s)+r_{1,2}}.\label{eq:recELadder}
\end{equation}
Again, the computation starts with the impedance where the electrical ladder only has one generation,
\begin{equation}
    G_1(s)=r_{1,1}+\cfrac{1}{\cfrac{1}{r_{1,2}}+c_1s}=\frac{r_{1,1}r_{1,2}c_1s+r_{1,1}+r_{1,2}}{r_{1,2}c_1s+1}.\label{eq:recELadder1}
\end{equation}
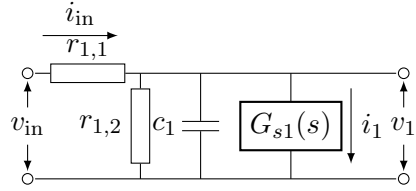
\begin{figure}
    \centering
    \begin{tikzpicture}[circuit ee IEC]
        \draw (0,0) circle (2pt);
        \draw (0.1,0) to [resistor] node [midway,above=1pt] {$r_{1,1}$} (1.5,0);
        \draw (1.5,0) to [resistor] node [midway,left=3pt] {$r_{1,2}$} (1.5,-1.4);
        \draw (2.3,0) to [capacitor] node [midway,left=5pt] {$c_1$} (2.3,-1.4);
        \node[draw,outer sep=0pt,thick] (G1) at (3.5,-0.7) {$G_{s1}(s)$};
        \draw[thin] (G1.north) -- (3.5,0);
        \draw[thin] (G1.south) -- (3.5,-1.4);
        \draw[thin, -latex] (4.3,-0.2) -- node [midway,right=0.1] {$i_1$} (4.3,-1.2);
        \node at (5,-0.7) (v1) {$v_1$};
        \draw[thin] (1.5,0) -- (4.9,0);
        \draw (0,-1.4) circle (2pt);
        \draw[thin] (0.1,-1.4) -- (4.9,-1.4);
        \node at (0,-0.7) (vi) {$v_{\text{in}}$};
        \draw[thin, -latex] (vi.north) -- (0,-0.1);
        \draw[thin, -latex] (vi.south) -- (0,-1.3);
        \draw (5,0) circle (2pt);
        \draw (5,-1.4) circle (2pt);
        \draw[thin, -latex] (v1.north) -- (5,-0.1);
        \draw[thin, -latex] (v1.south) -- (5,-1.3);
        \draw[thin, -latex] (0.2,0.5) -- node [midway,above=0.1] {$i_{\text{in}}$} (1.2,0.5);
    \end{tikzpicture}
    \caption{An illustration for obtaining the electrical ladder's recurrence formula}
    \label{fig:recELadder}
\end{figure}

For the mechanical ladder network in \Cref{fig:mLadder}, its recurrence formula is less straightforward to evaluate since the masses are included. The sketch for the derivation is shown in \Cref{fig:recMLadder}, where $x_1$ denotes the distance between $m_1$ and $m_2$, while $x_2$ indicates the length of the sub-network between $m_2$ and $m_{\text{last}}$. Here, we assume the transfer function for that sub-network is available as
\begin{equation*}
    G_{s1}(s)=\frac{X_2(s)}{F_{\text{PID}_1}(s)},
\end{equation*}
where the force $f_{\text{PID}_1}$ is exerted by the $\text{PID}_1$ controller on $m_2$. Therefore,
\begin{equation*}
    F_{\text{PID}_1}(s)= \left( k_{p1}+\frac{k_{i1}}{s}+k_{d1}s \right) X_1(s)=K_1(s)X_1(s).
\end{equation*}
Hence, we now know that
\begin{equation}
    G_{s1}(s)=\frac{X_2(s)}{K_1(s)X_1(s)}.
    \label{eq:recMLadderDer1}
\end{equation}
Our goal is to derive the transfer function for the entire network
\begin{equation}
    G_r(s)=\frac{X(s)}{F(s)}=\frac{X_1(s)+X_2(s)}{F(s)},
    \label{eq:recMLadderDer2}
\end{equation}
given $G_{s1}(s)$. Due to our assumption that $m_{\text{last}}$ always moves at a constant speed as mentioned in \Cref{sec:assum}, we have the following relation from Newton's second law of motion,
\begin{equation}
    m_1s^2(X_1(s)+X_2(s))=F(s)-K_1(s)X_1(s)-b_1s(X_1(s)+X_2(s)).\label{eq:recMLadderDer3}
\end{equation}
Combining the above three Eqs.~\cref{eq:recMLadderDer1} to \cref{eq:recMLadderDer3} together, we can obtain the mechanical ladder's recurrence formula
\begin{equation}
    G_r(s)=\frac{G_{s1}(s)K_1(s)+1}{(m_1s^2+b_1s)(G_{s1}(s)K_1(s)+1)+K_1(s)}.\label{eq:recMLadder}
\end{equation}
The computation starts with the transfer function where the mechanical ladder only has one generation,
\begin{equation}
    G_1(s)=\frac{1}{m_1s^2+b_1s+K_1(s)}.\label{eq:recMLadder1}
\end{equation}
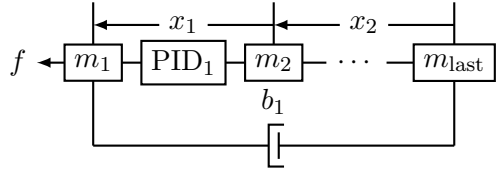
\begin{figure}
    \centering
    \begin{tikzpicture}
    \tikzstyle{spring}=[thick,decorate,decoration={zigzag,pre length=0.3cm,post length=0.3cm,segment length=6}]
        \tikzstyle{damper}=[thick,decoration={markings,mark connection node=dmp,mark=at position 0.5 with 
        {
            \node (dmp) [thick,inner sep=0pt,transform shape,rotate=-90,minimum width=15pt,minimum height=3pt,draw=none] {};
            \draw [thick] ($(dmp.north east)+(2pt,0)$) -- (dmp.south east) -- (dmp.south west) -- ($(dmp.north west)+(2pt,0)$);
            \draw [thick] ($(dmp.north)+(0,-5pt)$) -- ($(dmp.north)+(0,5pt)$);
        }
        },decorate]
        \node at (0,0) (f) {$f$};
        \node[draw,outer sep=0pt,thick] (m1) at (1,0) {$m_1$};
        \node[draw,outer sep=0pt,thick] (c1) at (2.2,0) {$\text{PID}_1$};
        \node[draw,outer sep=0pt,thick] (m2) at (3.4,0) {$m_2$};
        \node[outer sep=0pt,thick] at (4.5,0) {$\cdots$};
        \node[draw,outer sep=0pt,thick] (mL) at (5.8,0) {$m_{\text{last}}$};
        \draw[thick, -latex] (m1.west) -- (f.east);
        \draw[thick] (m1.east) -- (c1.west);
        \draw[thick] (c1.east) -- (m2.west);
        \draw[thick] (m2.east) -- (4.1,0);
        \draw[thick] (4.9,0) -- (mL.west);
        \draw[damper] (1,-1.1) -- (5.8,-1.1) node [midway,above=8pt] {$b_1$};
        \draw[thick] (m1.south) -- (1,-1.1);
        \draw[thick] (mL.south) -- (5.8,-1.1);
        \draw[thick] (m1.north) -- (1,0.8);
        \draw[thick] (m2.north) -- (3.4,0.8);
        \draw[thick] (mL.north) -- (5.8,0.8);
        \node[outer sep=0pt,thick] (x1) at (2.2,0.5) {$x_1$};
        \node[outer sep=0pt,thick] (x2) at (4.6,0.5) {$x_2$};
        \draw[thick,-latex] (x1.west) -- (1,0.5);
        \draw[thick] (x1.east) -- (3.4,0.5);
        \draw[thick,-latex] (x2.west) -- (3.4,0.5);
        \draw[thick] (x2.east) -- (5.8,0.5);
    \end{tikzpicture}
    \caption{An illustration for obtaining the mechanical ladder network's recurrence formula}
    \label{fig:recMLadder}
\end{figure}

\subsection{Frequency response}
\label{sec:finModNum}
As discussed at the beginning of \Cref{sec:finMod}, it is natural to see that by repeatedly using the recurrence formula, a finite network's frequency response can be constructed starting from the deepest generation all the way back to the first generation. For example, we want to evaluate the frequency response of a three-generation tree network. The first thing is to compute that of the one-generation sub-tree between $x_{3,1}$ and $x_{\text{last}}$. (See \Cref{fig:tree}.) To do that, we need to use the one-generation formula $G_1(s)$ from \cref{eq:recTree1} where $k_{1,1}$ and $b_{1,1}$ should be substituted with the values of $k_{3,1}$ and $b_{3,1}$. Then, we need to repeat the same computation three more times for all the other one-generation sub-trees between the third generation and $x_{\text{last}}$. Next, we go to the second generation and use the recurrence formula~\cref{eq:recTree} for the two-generation sub-tree between $x_{2,1}$ and $x_{\text{last}}$. At this occurrence, $k_{1,1}$ and $b_{1,1}$ in ~\cref{eq:recTree} should be replaced by the values of $k_{2,1}$ and $b_{2,1}$. In addition, two of the four frequency response which we just computed are put at the positions of $G_{s1}(s)$ and $G_{s2}(s)$ accordingly. Then, we repeat this again for the other two-generation sub-tree between $x_{2,2}$ and $x_{\text{last}}$. Finally, we use the recurrence formula~\cref{eq:recTree} once more to compute the frequency response for the entire network between $x_{1,1}$ and $x_{\text{last}}$.

The above example shows that such a backward computation from the deepest generation to the first generation is challenging to code in a systematic manner for different networks. Furthermore, it depends on users to correctly put different components' constants at their respective iterations throughout the calculation which is unwieldy for large networks and is thus error-prone. To overcome these challenges, we propose a recursive algorithm listed in \Cref{alg:numFin}.

\begin{algorithm}
    \caption{Pseudocode of our modeling algorithm for finite networks' frequency response. It computes the frequency response \texttt{G} at the angular frequency \texttt{w} for a finite network with \texttt{nG} number of generations given its damage case \texttt{(l,e)} and the undamaged constants \texttt{undCst}.}
    \label{alg:numFin}
    \begin{algorithmic}[1]
    \STATE{\texttt{\textbf{function} G~=~freqFin(l,e,undCst,w,nG)}}
    \STATE{\texttt{s~=~i*w;}}
    \STATE{\texttt{[l1,e1,lS,eS]~=~partition(l,e);}}
    \STATE{\texttt{g1Cst~=~getG1Cst(l1,e1,undCst);}}
    \IF{\texttt{nG~==~1}}
    \STATE{\texttt{G~=~G1(g1Cst,s);}}
    \ELSE
    \STATE{\texttt{nG~=~nG-1;}}
    \FOR{\texttt{idx from 1 to nS}}
    \STATE{\texttt{GS[idx]~=~freqFin(lS[idx],eS[idx],undCst,w,nG);}}
    \ENDFOR
    \STATE{\texttt{G~=~Gr(g1Cst,GS,s);}}
    \ENDIF
    \end{algorithmic}
\end{algorithm}

The algorithm first converts the angular frequency \texttt{w} to the Laplace variable \texttt{s} and then it employs the \texttt{partition()} function to partition the entire network's damage case \texttt{(l,e)} into two parts. One part is the damage case of the first-generation components \texttt{(l1,e1)}. The other part is a collection of damage cases, \texttt{(lS[idx],eS[idx])}, where each one concerns the \texttt{idx}-th sub-network between the second generation and the last. A concrete example for this \texttt{partition()} function was provided in our previous paper \cite{ni2020damage}. Next, the algorithm evaluates the first-generation components' constants \texttt{g1Cst} given the corresponding damage case \texttt{(l1,e1)} and the undamaged constants \texttt{undCst} per the descriptions in \Cref{sec:assum}. Then, the algorithm splits into two branches determined by the \texttt{if}-condition whose criterion is whether the input argument for the number of generations \texttt{nG} equals to one. If so, the returned value \texttt{G} is the result of those one-generation frequency response, that is Eqs.~\cref{eq:recTree1}, \cref{eq:recELadder1} and \cref{eq:recMLadder1} in the examples. Otherwise, the number of generations \texttt{nG} is decreased by one and is then used to recursively call the algorithm itself for each of the sub-networks whose frequency response \texttt{Gs[idx]} is calculated during those recursive calls . As a result, the tree network needs two recursive calls, while both the electrical and the mechanical ladders only require one. In the end, the returned frequency response \texttt{G} is computed by the \texttt{Gr()} function according to the recurrence formulas, such as Eqs.~\cref{eq:recTree}, \cref{eq:recELadder} and \cref{eq:recMLadder} in the examples.

Due to the recursive nature of the proposed algorithm, it is still doing the same backward computation which is handled automatically. As opposed to the manual computation mentioned above, while coding this algorithm, we merely need to focus on the first generation, which is much less cumbersome for large networks. Additionally, the algorithm is modular so that all of its parts, such as the \texttt{G1()} and \texttt{Gr()} functions are interchangeable for different networks. 

\begin{figure}
    \centering
    \includegraphics[width=.75\textwidth]{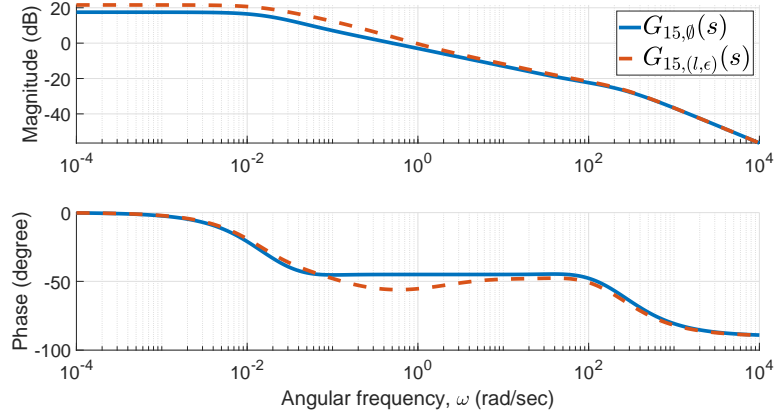}
    \caption{Frequency response for 15-generation tree networks. The blue curve is for the undamaged case, $G_{15,\varnothing}(i\omega)$. The red dashed curve is for a damage case, $G_{15,(\boldsymbol{l},\boldsymbol{\epsilon})}(i\omega)$, where $(\boldsymbol{l},\boldsymbol{\epsilon})=([k_{2,1},k_{2,2},b_{3,1}],[0.1,0.2,0.3])$. The undamaged constants are $k=2N/m$ and $b=1Ns/m$.}
    \label{fig:treeNumFin}
\end{figure}

\begin{figure}
    \centering
    \includegraphics[width=.75\textwidth]{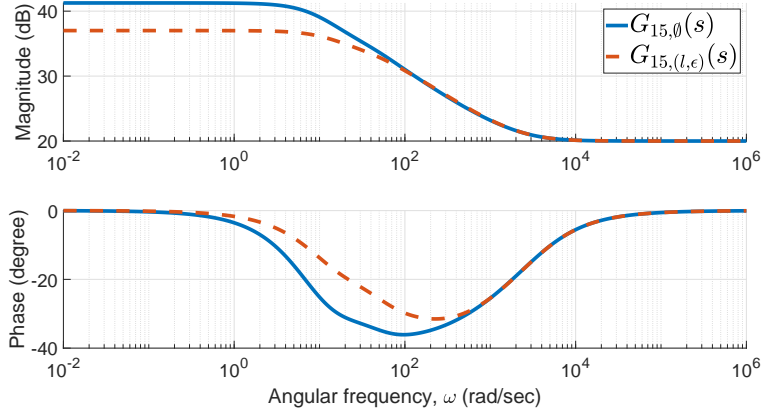}
    \caption{Numerical input impedance for 15-generation electrical ladder networks. The blue curve is for the undamaged case, $G_{15,\varnothing}(i\omega)$. The red dashed curve is for a damage case, $G_{15,(\boldsymbol{l},\boldsymbol{\epsilon})}(i\omega)$, where $(\boldsymbol{l},\boldsymbol{\epsilon})=([r_{2,2}],[0.1])$. The undamaged constants are $r_1=10\Omega$, $r_2=1k\Omega$ and $c=100\mu F$.}
    \label{fig:eLadderNumFin}
\end{figure}

\begin{figure}
    \centering
    \includegraphics[width=.75\textwidth]{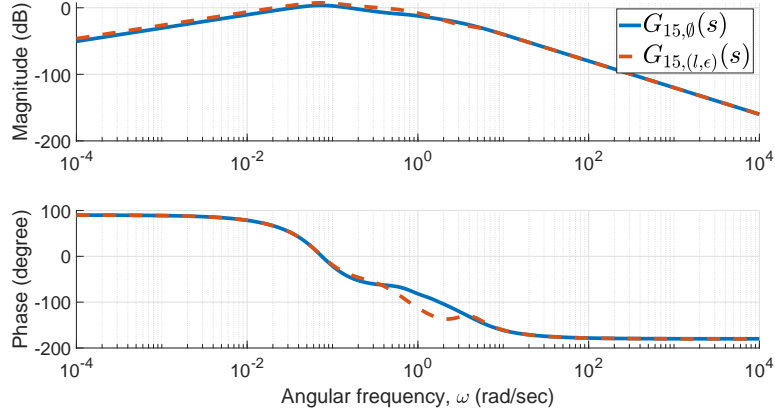}
    \caption{Frequency response for 15-generation mechanical ladder networks. The blue curve is for the undamaged case, $G_{15,\varnothing}(i\omega)$. The red dashed curve is for a damage case, $G_{15,(\boldsymbol{l},\boldsymbol{\epsilon})}(i\omega)$, where $(\boldsymbol{l},\boldsymbol{\epsilon})=([k_{p2},k_{i2},k_{d2}],[0.1,0.1,0.1])$. The undamaged constants are $k_p=10N/m$, $k_i=0.5N/ms$, $k_d=2Ns/m$ and $b=1Ns/m$.}
    \label{fig:mLadderNumFin}
\end{figure}

Note that the algorithm listed in \Cref{alg:numFin}, as all the algorithms proposed in this paper, works for both damaged and undamaged networks. When the input argument \texttt{l} is an empty list, the algorithm should return the undamaged frequency response. \Cref{fig:treeNumFin} to \Cref{fig:mLadderNumFin} illustrate the resultant frequency response for the finite version of those three example networks. Note that the undamaged constants labeled in the captions of those figures are used consistently throughout this paper. However, we tested the results for a variety of undamaged constants, which have no qualitative differences from the ones included in this paper.

\subsection{Transfer functions}
\label{sec:finModAna}
Determining the transfer function for a network utilizes the same recursive algorithm mentioned above with a modification on recurrence formulas so that their computations can be performed purely on the coefficients and thus are independent of the Laplace variable $s$. That modification is based on the following two observations.
\begin{enumerate}
    \item The one-generation frequency responses, $G_1(s)$, are rational expressions in closed-form.
    \item If the sub-networks' frequency responses, $G_s(s)$, are rational expressions, so are the results returned by the recurrence formulas.
\end{enumerate}
Combining the above two observations together, we can conclude that all frequency responses for finite networks that satisfy the assumptions from (A-1) to (A-4) are rational expressions, where both numerator and denominator are polynomials in $s$. In other words, all finite networks' transfer functions $G(s)$ are
\begin{equation*}
    G(s)=\frac{N(s)}{D(s)}=\frac{c_N\mu_N(s)}{c_D\mu_D(s)},
\end{equation*}
where $c_N$, $c_D$ are coefficient vectors, and $\mu_N(s)$, $\mu_D(s)$ are the corresponding monomial bases for the polynomials in the numerator $N(s)$ and the denominator $D(s)$. Therefore, using the fact that the multiplication of two polynomials is equivalent to the convolution of their coefficient vectors, we can rewrite the recurrence formulas so that their computations can be directly conducted on the coefficients.

We arrange the coefficients from the highest order to the lowest. Then, the one-generation frequency responses, $G_1(s)$, and the recurrence formulas, $G_r(s)$, for those three example networks can be recast as follows. For the tree, from Eq.~\cref{eq:recTree1}, we know that the coefficient vectors for $G_1(s)$ are
\begin{align}
    c_{N1}&=\begin{bmatrix}1\end{bmatrix},\label{eq:recTreeCoeff1_1}\\
    c_{D1}&=\begin{bmatrix}b_{1,1}&k_{1,1}\end{bmatrix}.\label{eq:recTreeCoeff1_2}
\end{align}
Next, we assume the tree's two sub-networks' analytical frequency responses are also rational expressions. That is,
\begin{align}
    G_{s1}(s)&=\frac{N_{s1}(s)}{D_{s1}(s)}=\frac{c_{Ns1}\mu_{Ns1}(s)}{c_{Ds1}\mu_{Ds1}(s)},\label{eq:finModAnaTreeR1}\\
    G_{s2}(s)&=\frac{N_{s2}(s)}{D_{s2}(s)}=\frac{c_{Ns2}\mu_{Ns2}(s)}{c_{Ds2}\mu_{Ds2}(s)}.\label{eq:finModAnaTreeR2}
\end{align}
Substituting~\cref{eq:finModAnaTreeR1} and \cref{eq:finModAnaTreeR2} into the recurrence formula, $G_r(s)$, Eq.~\cref{eq:recTree}, we then have
\begin{equation*}
    G_r(s)=\frac{N_r(s)}{D_r(s)}=\frac{k_{1,1}b_{1,1}sN_{s1}N_{s2}+k_{1,1}N_{s1}D_{s2}+b_{1,1}sN_{s2}D_{s1}+D_{s1}D_{s2}}{k_{1,1}b_{1,1}s(N_{s1}D_{s2}+N_{s2}D_{s1})+(k_{1,1}+b_{1,1}s)D_{s1}D_{s2}}.
\end{equation*}
Then, by utilizing convolutions, we can rewrite the recurrence formula $G_r(s)$ into the version where only its coefficient vectors are concerned. That is,
\begin{align}
    c_{Nr}&=\begin{bmatrix}k_{1,1}b_{1,1}&0\end{bmatrix}*c_{Ns1}*c_{Ns2}+\begin{bmatrix}k_{1,1}\end{bmatrix}*c_{Ns1}*c_{Ds2}+\begin{bmatrix}b_{1,1}&0\end{bmatrix}*c_{Ns2}*c_{Ds1}+c_{Ds1}*c_{Ds2},\label{eq:recTreeCoeff1}\\
    c_{Dr}&=\begin{bmatrix}k_{1,1}b_{1,1}&0\end{bmatrix}*(c_{Ns1}*c_{Ds2}+c_{Ns2}*c_{Ds1})+\begin{bmatrix}b_{1,1}&k_{1,1}\end{bmatrix}*c_{Ds1}*c_{Ds2},\label{eq:recTreeCoeff2}
\end{align}
where the operator $*$ denotes the convolution between two vectors. Note that all additions above can be carried out between two vectors with different lengths, whose definition agrees with the addition between two polynomials. For instance, we could have
\begin{equation*}
    \begin{bmatrix}1&2&3\end{bmatrix}+\begin{bmatrix}4&5\end{bmatrix}=\begin{bmatrix}1&6&8\end{bmatrix}.
\end{equation*}

For the electrical ladder, following the similar procedure, we can obtain from Eq.~\cref{eq:recELadder1} that
\begin{align*}
    c_{N1}&=\begin{bmatrix}r_{1,1}r_{1,2}c_1&r_{1,1}+r_{1,2}\end{bmatrix},\\
    c_{D1}&=\begin{bmatrix}r_{1,2}c_1&1\end{bmatrix}.
\end{align*}
From Eq.~\cref{eq:recELadder}, we can get that the revised recurrence formulas are
\begin{align*}
    c_{Nr}&=\begin{bmatrix}r_{1,1}r_{1,2}c_1&r_{1,1}+r_{1,2}\end{bmatrix}*c_{Ns1}+r_{1,1}r_{1,2}c_{Ds1},\\
    c_{Dr}&=\begin{bmatrix}r_{1,2}c_1&1\end{bmatrix}*c_{Ns1}+r_{1,2}c_{Ds1}.
\end{align*}

For the mechanical ladder, from Eq.~\cref{eq:recMLadder1}, its one-generation coefficient vectors are
\begin{align*}
    c_{N1}&=\begin{bmatrix}1&0\end{bmatrix},\\
    c_{D1}&=\begin{bmatrix}m_1&b_1+k_{d1}&k_{p1}&k_{i1}\end{bmatrix}.
\end{align*}
From Eq.~\cref{eq:recMLadder}, its revised recurrence formula are
\begin{align*}
    c_{Nr}&=\begin{bmatrix}k_{d1}&k_{p1}&k_{i1}\end{bmatrix}*c_{Ns1}+\begin{bmatrix}1&0\end{bmatrix}*c_{Ds1},\\
    c_{Dr}&=\begin{bmatrix}m_1&b_1&0\end{bmatrix}*\begin{bmatrix}k_{d1}&k_{p1}&k_{i1}\end{bmatrix}*c_{Ns1}+\begin{bmatrix}m_1&b_1+k_{d1}&k_{p1}&k_{i1}\end{bmatrix}*c_{Ds1}.
\end{align*}

After transforming $G_1(s)$ and $G_r(s)$, we need to adjust that recursive algorithm accordingly, which is listed in \Cref{alg:anaFin}. The \texttt{C1()} function contains the computations of the coefficient vectors for one-generation networks, $c_{N1}$ and $c_{D1}$, such as Eqs.~\cref{eq:recTreeCoeff1_1} and \cref{eq:recTreeCoeff1_2} for the tree. The \texttt{Cr()} function includes the revised recurrence formulas, $c_{Nr}$ and $c_{Dr}$, \textit{e.g.} Eqs.~\cref{eq:recTreeCoeff1} and \cref{eq:recTreeCoeff2} for the tree. The \texttt{simplify()} function is optional which simplifies the coefficient vectors of the numerator and denominator, such as dividing all coefficients by a common value so that one of those coefficients equals one. Compared to the algorithm for frequency response listed in \Cref{alg:numFin}, the differences are that the algorithm no longer relies on some specific angular frequencies \texttt{w}, and it returns the coefficient vectors \texttt{cN} and \texttt{cD} of a network's transfer function $G(s)$ instead of its numerical value. Based on the returned coefficient vectors \texttt{cN} and \texttt{cD}, we can immediately tell the correlated monomial bases and thus know the expression of $G(s)$.

\begin{algorithm}
    \caption{Pseudocode of the algorithm for finite networks' transfer function. It computes the coefficient vectors \texttt{cN} and \texttt{cD} of an \texttt{nG}-generation network's transfer function given its damage case \texttt{(l,e)} and the undamaged constants \texttt{undCst}.}
    \label{alg:anaFin}
    \begin{algorithmic}[1]
    \STATE{\texttt{\textbf{function}~[cN,cD]~=~tranFin(l,e,undCst,nG)}}
    \STATE{\texttt{[l1,e1,lS,eS]~=~partition(l,e);}}
    \STATE{\texttt{g1Cst~=~getG1Cst(l1,e1,undCst);}}
    \IF{nG~==~1}
    \STATE{\texttt{[cN,cD]~=~C1(g1Cst);}}
    \ELSE
    \STATE{\texttt{nG~=~nG-1;}}
    \FOR{\texttt{idx from 1 to nS}}
    \STATE{\texttt{[cNS[idx],cDS[idx]]~=~tranFin(lS[idx],eS[idx],undCst,nG);}}
    \ENDFOR
    \STATE{\texttt{[cN,cD]~=~Cr(g1Cst,cNS,cDS);}}
    \STATE{\texttt{[cN,cD]~=~simplify(cN,cD);}}
    \ENDIF
    \end{algorithmic}
\end{algorithm}

For example, for a two-generation tree whose damage case is $(\boldsymbol{l},\boldsymbol{\epsilon})=([k_{2,1},k_{2,2}],[0.1,0.2])$, its transfer function is
\begin{equation*}
    G_{2,(\boldsymbol{l},\boldsymbol{\epsilon})}(s)=\frac{2s^2+4.8s+0.88}{s^3+6.6s^2+2.48s+0.16}.
\end{equation*}
For a four-generation electrical ladder whose damage case is $(\boldsymbol{l},\boldsymbol{\epsilon})=([r_{2,2},r_{3,2}],[0.1,0.1])$, its input impedance's analytical expression is
\begin{equation*}
    G_{4,(\boldsymbol{l},\boldsymbol{\epsilon})}(s)=\frac{s^4+7220s^3+1.6\times10^7s^2+1.2\times10^{10}s+1.6\times10^{12}}{0.1s^4+622s^3+1.1\times10^6s^2+5\times10^8s+2.4\times10^{10}}.
\end{equation*}
For a two-generation mechanical ladder whose damage case is
\[ (\boldsymbol{l},\boldsymbol{\epsilon})=([k_{p2},k_{i2},k_{d2}],[0.1,0.1,0.1],
\]
its transfer function is
\begin{equation*}
    G_{2,(\boldsymbol{l},\boldsymbol{\epsilon})}(s)=\frac{s^4+3.2s^3+11s^2+0.55s}{s^6+6.2s^5+26.6s^4+26.05s^3+11.25s^2+s+0.025}.
\end{equation*}

\section{Infinite networks}
\label{sec:infMod}
For an infinite network to be eligible for the modeling algorithms proposed in this section, it must fulfill two more assumptions except for the assumptions (A-1) to (A-4).
\begin{itemize}
    \item (A-5) The network has a finite number of damaged components.
    \item (A-6) The network's undamaged transfer function can be obtained.
\end{itemize}
Because the network is infinitely large, the assumption (A-5) equivalently requires that there exists a generation inside that network after which all sub-networks are undamaged. Then, that generation acts as the starting point for the same backward computation procedure which is used for finite networks in \Cref{sec:finMod}. In other words, an infinite network's undamaged transfer function plays the same role as the one-generation transfer function $G_1(s)$ does inside the finite networks' modeling algorithm, with which the recurrence formula $G_r(s)$ starts gradually building up the entire network. As a result, we need to compute the undamaged transfer function for an infinite network at first, which is reviewed in \Cref{sec:infModUnd}. That is the reason why the assumption (A-6) is also necessary. The undamaged frequency response is then utilized by the similar recursive algorithms which are modified accordingly for infinite networks as illustrated in \Cref{sec:infModD}. Furthermore, for the computing the transfer function in \Cref{sec:infModDAna}, we still use the convolution to convert the recurrence formulas. However, the difference is that for infinite networks, the polynomials involved no longer only consist of the integer orders of $s$. In contrast, the fractional orders of $s$ or even some irrational functions of $s$ naturally become the elements for those polynomials, which is a major characterization of infinite networks' dynamics as opposed to the finite ones'.

\subsection{Undamaged networks}
\label{sec:infModUnd}
The contents in this section for undamaged infinite networks are from \cite{mayes2012reduction}. For purposes of completeness and coherency of presentation, we briefly summarize the main features of those results. The main idea is taking advantage of an infinite undamaged network's self-repeating nature and re-writing its recurrence formula into a special form where its transfer function can be solved.

For an undamaged tree network, when it is infinitely large, both sub-networks at the second generation are identical, which are correspondingly also identical to the entire tree. In addition, both components at the first generation are intact, which means $k_{1,1}=k$ and $b_{1,1}=b$. Hence, in this case, if we use $G_{\infty,\varnothing}(s)$ to denote the transfer function of the undamaged infinite tree, its recurrence formula~\cref{eq:recTree} becomes
\begin{equation*}
    G_{\infty,\varnothing}(s)=\cfrac{1}{\cfrac{1}{\cfrac{1}{k}+G_{\infty,\varnothing}(s)}+\cfrac{1}{\cfrac{1}{bs}+G_{\infty,\varnothing}(s)}}=\frac{kbsG_{\infty,\varnothing}^2(s)+(k+bs)G_{\infty,\varnothing}(s)+1}{2kbsG_{\infty,\varnothing}(s)+k+bs},
\end{equation*}
where $G_{\infty,\varnothing}(s)$ is the only unknown. As a result, the undamaged infinite tree's transfer function can be solved in closed-form as
\begin{equation}
    G_{\infty,\varnothing}(s)=\frac{1}{\sqrt{kbs}}.
    \label{eq:treeUndInf}
\end{equation}

For the undamaged infinite electrical ladder network, its recurrence formula~\cref{eq:recELadder} can be re-written as
\begin{equation*}
    G_{\infty,\varnothing}(s)=\frac{(r_1r_2cs+r_1+r_2)G_{\infty,\varnothing}(s)+r_1r_2}{(r_2cs+1)G_{\infty,\varnothing}(s)+r_2},
\end{equation*}
which leads to the fact that its input impedance is
\begin{equation}
    G_{\infty,\varnothing}(s)=\cfrac{s+\cfrac{1}{r_2c}+\sqrt{s^2+\cfrac{2r_1+4r_2}{r_1r_2c}s+\cfrac{r_1+4r_2}{r_1r_2^2c^2}}}{\cfrac{2}{r_1}s+\cfrac{2}{r_1r_2c}}.
    \label{eq:eLadderUndInf}
\end{equation}
Similarly, for the undamaged mechanical ladder network, its recurrence formula~\cref{eq:recMLadder} can be re-written as
\begin{equation*}
    G_{\infty,\varnothing}(s)=\frac{G_{\infty,\varnothing}(s)K(s)+1}{(ms^2+bs)(G_{\infty,\varnothing}(s)K(s)+1)+K(s)},
\end{equation*}
where $K(s)=k_p+k_i/s+k_ds$, which yields that
\begin{equation}
    G_{\infty,\varnothing}(s)\nonumber=\frac{-ms^2-bs+A(s)}{2[mk_ds^3+(mk_p+bk_d)s^2+(mk_i+bk_p)s+bk_i]},\label{eq:mLadderUndInf}
\end{equation}
where
\begin{equation}
    A(s)=[m^2s^4+(2mb+4mk_d)s^3+(b^2+4mk_p+4bk_d)s^2+4(mk_i+bk_p)s+4bk_i]^{\frac{1}{2}}.\label{eq:mLadderUndInfA}
\end{equation}

\begin{figure}
    \centering
    \includegraphics[width=.75\textwidth]{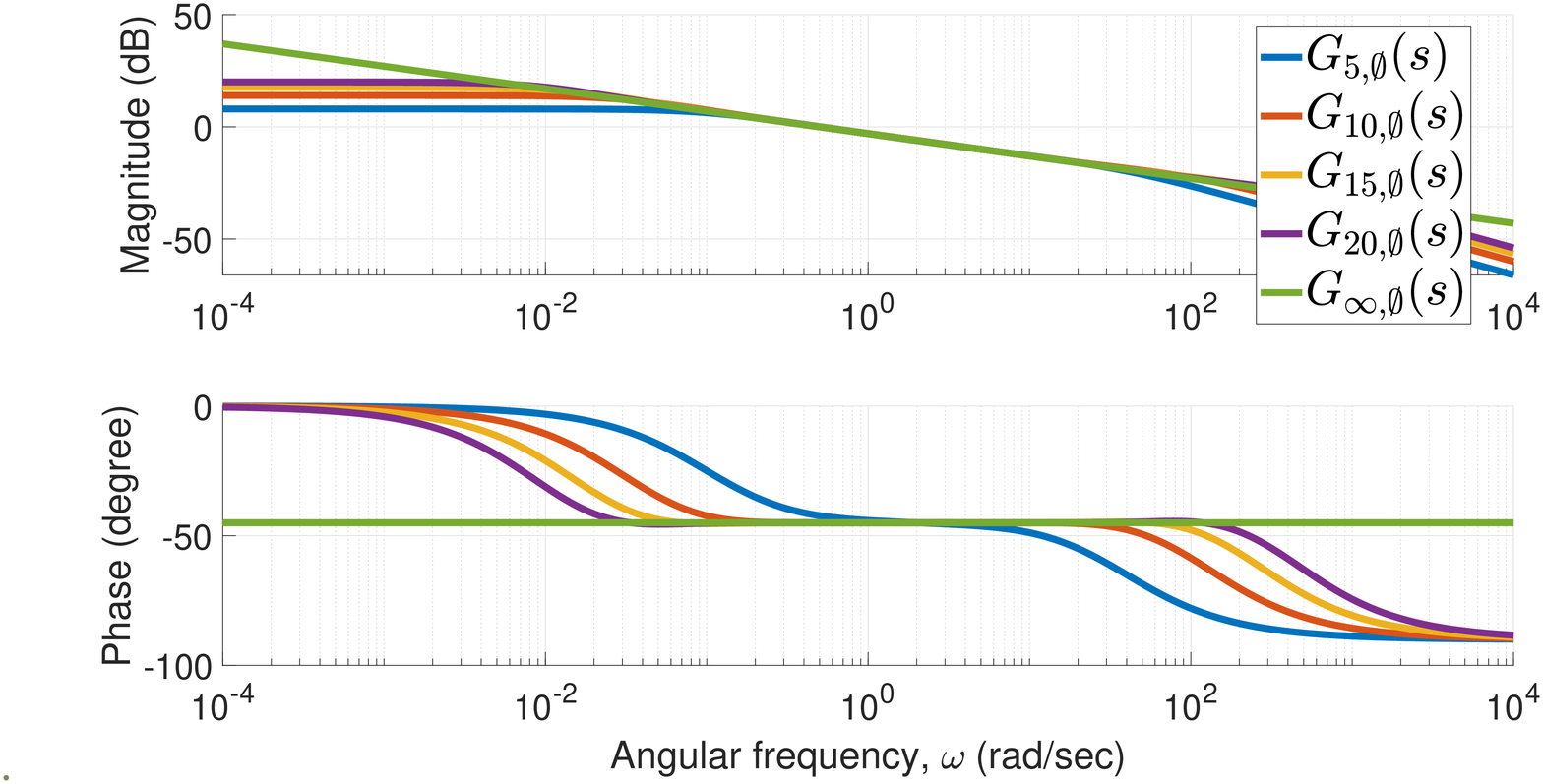}
    \caption{Frequency response of undamaged $g$-generation tree networks, $G_{g,\varnothing}(i\omega)$, converge to that of the infinite version, $G_{\infty,\varnothing}(i\omega)$.}
    \label{fig:treeUndInf}
\end{figure}

\begin{figure}
    \centering
    \includegraphics[width=.75\textwidth]{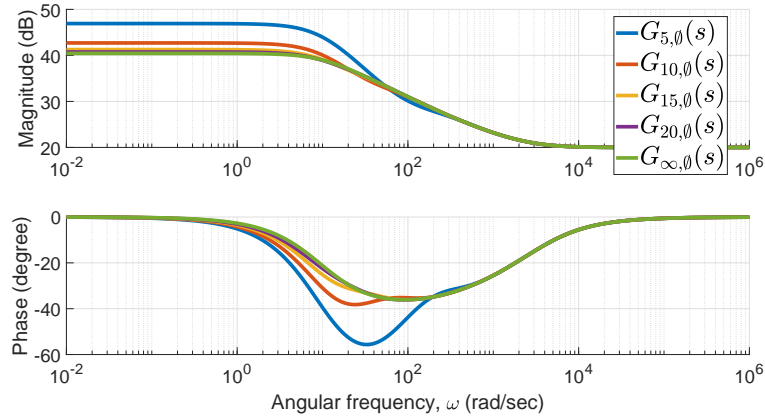}
    \caption{Input impedance of undamaged $g$-generation electrical ladder networks, $G_{g,\varnothing}(i\omega)$, converges to that of the infinite version, $G_{\infty,\varnothing}(i\omega)$.}
    \label{fig:eLadderUndInf}
\end{figure}

\begin{figure}
    \centering
    \includegraphics[width=.75\textwidth]{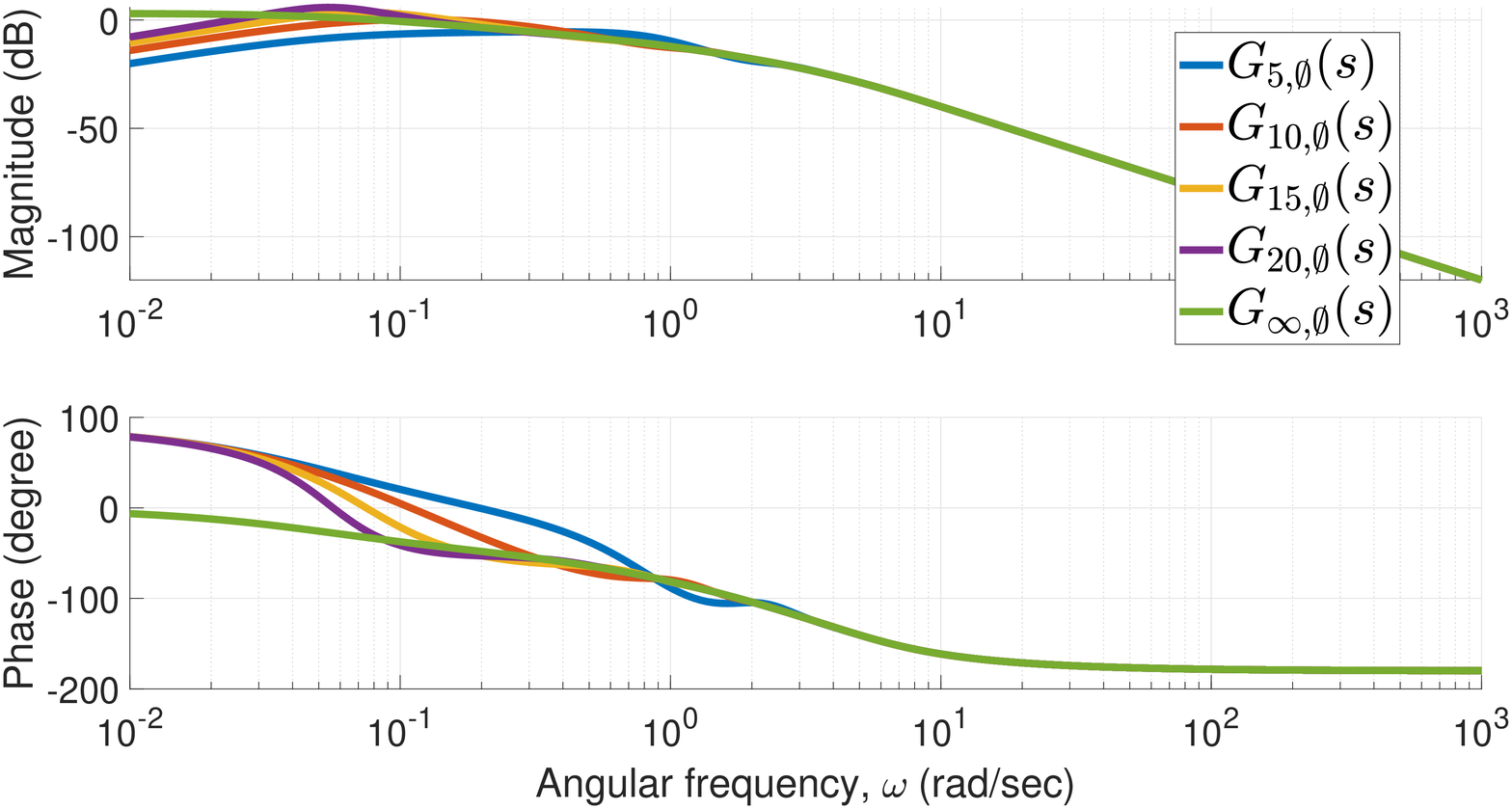}
    \caption{Frequency response of undamaged $g$-generation mechanical ladder networks, $G_{g,\varnothing}(i\omega)$, converge to that of the infinite version, $G_{\infty,\varnothing}(i\omega)$.}
    \label{fig:mLadderUndInf}
\end{figure}

\Cref{fig:treeUndInf} to \Cref{fig:mLadderUndInf} illustrate the undamaged frequency responses of those three example networks, from which we can confirm that the results for finite networks, $G_{g,\varnothing}(i\omega)$, converge to that for infinite ones, $G_{\infty,\varnothing}(i\omega)$, as the number of generations $g$ goes to infinity.

\subsection{Damaged networks}
\label{sec:infModD}
This section reveals how to compute an infinite damaged network's frequency response and transfer function using its undamaged ones obtained by using the procedure in \Cref{sec:infModUnd}. The methods for both frequency response and transfer functions are similar to their counterparts for finite networks in \Cref{sec:finMod} with some necessary modifications.

\subsubsection{Frequency response}
As discussed at the beginning of \Cref{sec:infMod}, our modeling procedures handle an infinite network as if it is finite where the corresponding undamaged frequency response, $G_{\infty,\varnothing}(i\omega)$, is regarded as the one-generation frequency response, $G_1(i\omega)$, which initiates the entire backward modeling procedure. Considering an infinite network to be finite is justified by assumption (A-5) which implies that if we go deep inside an infinite network, we would eventually encounter an infinite undamaged sub-network which is then treated as a one-generation sub-network since its undamaged frequency response is already known. Therefore, as listed in \Cref{alg:numInf}, the algorithm for infinite networks' frequency response is modified accordingly from its counterpart for finite networks.

\begin{algorithm}
    \caption{Pseudocode of our modeling algorithm for infinite networks' frequency response. It computes the frequency response \texttt{G} at the angular frequency \texttt{w} for an infinite network given its damage case \texttt{(l,e)} and the undamaged constants \texttt{undCst}.}
    \label{alg:numInf}
    \begin{algorithmic}[1]
    \STATE{\texttt{\textbf{function}~G~=~freqInf(l,e,undCst,w)}}
    \STATE{\texttt{s~=~i*w;}}
    \IF{\texttt{isEmpty(l)}}
    \STATE{\texttt{G~=~GUnd(undCst,s);}}
    \ELSE
    \STATE{\texttt{[l1,e1,lS,eS]~=~partition(l,e);}}
    \FOR{\texttt{idx from 1 to nS}}
    \STATE{\texttt{GS[idx]~=~freqInf(lS[idx],eS[idx],undCst,w);}}
    \ENDFOR
    \STATE{\texttt{g1Cst~=~getG1Cst(l1,e1,undCst);}}
    \STATE{\texttt{G~=~Gr(g1Cst,GS,s);}}
    \ENDIF
    \end{algorithmic}
\end{algorithm}

\begin{figure}
    \centering
    \includegraphics[width=.75\textwidth]{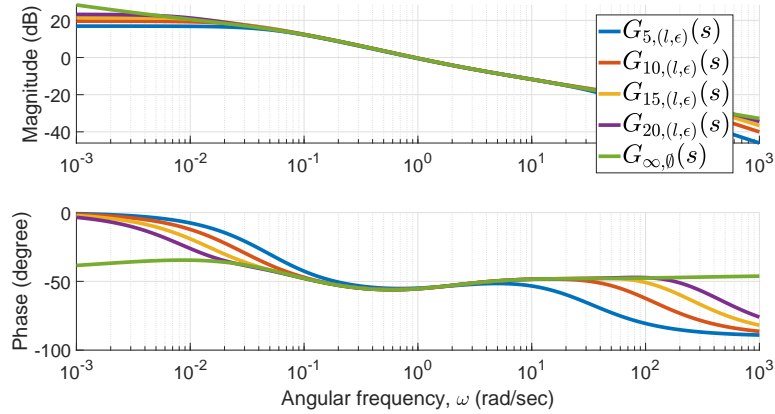}
    \caption{Frequency response of damaged $g$-generation tree networks, $G_{g,(\boldsymbol{l},\boldsymbol{\epsilon})}(i\omega)$, converge to that of the infinite version, $G_{\infty,(\boldsymbol{l},\boldsymbol{\epsilon})}(i\omega)$ where the damage case $(\boldsymbol{l},\boldsymbol{\epsilon})=([k_{2,1},k_{2,2},b_{3,1}],[0.1,0.2,0.3])$.}
    \label{fig:treeNumInf}
\end{figure}

\begin{figure}
    \centering
    \includegraphics[width=.75\textwidth]{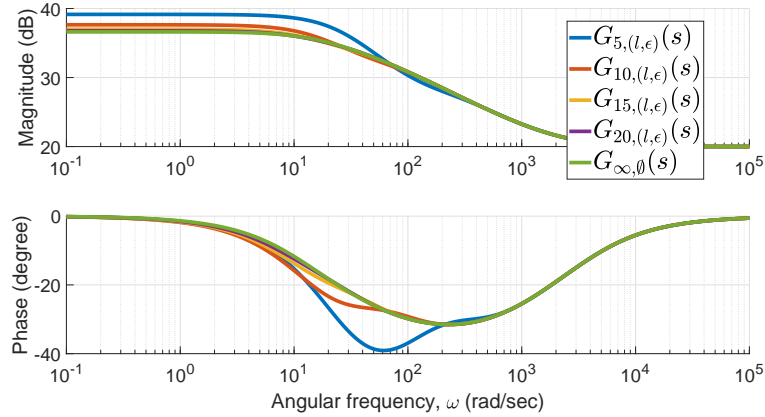}
    \caption{Input impedance of damaged $g$-generation electrical ladder networks, $G_{g,(\boldsymbol{l},\boldsymbol{\epsilon})}(i\omega)$, converges to that of the infinite version, $G_{\infty,(\boldsymbol{l},\boldsymbol{\epsilon})}(i\omega)$ where the damage case $(\boldsymbol{l},\boldsymbol{\epsilon})=([r_{2,2}],[0.1])$.}
    \label{fig:eLadderNumInf}
\end{figure}

\begin{figure}
    \centering
    \includegraphics[width=.75\textwidth]{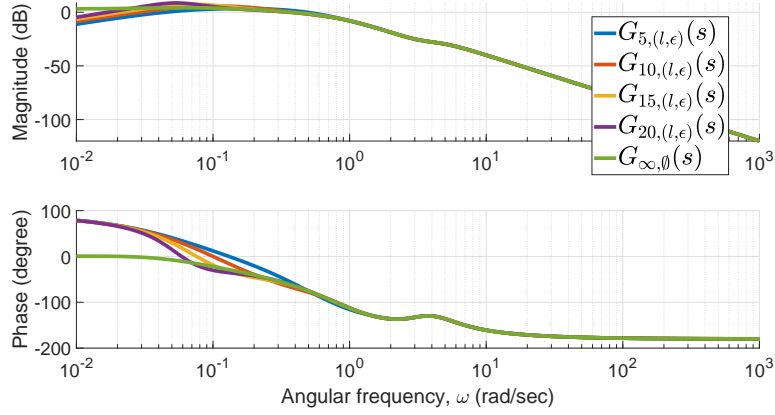}
    \caption{Frequency response of damaged $g$-generation mechanical ladder networks, $G_{g,(\boldsymbol{l},\boldsymbol{\epsilon})}(i\omega)$, converge to that of the infinite version, $G_{\infty,(\boldsymbol{l},\boldsymbol{\epsilon})}(i\omega)$ where the damage case $(\boldsymbol{l},\boldsymbol{\epsilon})=([k_{p2},k_{i2},k_{d2}],[0.1,0.1,0.1])$.}
    \label{fig:mLadderNumInf}
\end{figure}

Compared to the frequency response modeling algorithm for finite networks in \Cref{alg:numFin}, the main difference is that the criterion of the \texttt{if}-condition is revised to whether the network in query is undamaged, which is characterized by an empty list of damaged components, the input argument \texttt{l}. If so, the result is returned by the \texttt{GUnd()} function which computes the undamaged frequency response, such as Eqs.~\cref{eq:treeUndInf}, \cref{eq:eLadderUndInf} and \cref{eq:mLadderUndInf} for those three example networks. Otherwise, the algorithm follows exactly the same steps as those for finite networks. Especially, the computations in the \texttt{Gr()} function follow the recurrence formula, \textit{e.g.}, Eqs.~\cref{eq:recTree}, \cref{eq:recELadder} and \cref{eq:recMLadder}. \Cref{alg:numInf} affirms that the computation is again backward from the generation where all the sub-networks are undamaged all the way to the very first generation. \Cref{fig:treeNumInf} to \Cref{fig:mLadderNumInf} show the convergence between finite damaged networks and their infinite versions as their number of generation increases, where the infinite frequency responses, $G_{\infty,(\boldsymbol{l},\boldsymbol{\epsilon})}(i\omega)$, are computed by using \Cref{alg:numInf}.

\subsubsection{Transfer function}
\label{sec:infModDAna}
In \Cref{sec:finModAna} for the finite networks' transfer function, we extract the coefficient vectors from the one-generation transfer functions, $G_1(s)$, and use them to trigger the entire modeling procedure where the recurrence formulas are further adjusted by using convolutions so that pure coefficients computation is possible. As mentioned previously, the role of $G_1(s)$ is replaced by $G_{\infty,\varnothing}(s)$ for computing an infinite network's transfer function. As a result, it is clear that we need to extract the coefficients from the undamaged frequency response for infinite networks $G_{\infty,\varnothing}(s)$ in a similar manner. However, there exists a challenge that $G_{\infty,\varnothing}(s)$ is no longer a rational expression as opposed to $G_1(s)$, which is a common phenomenon for infinite self-similar networks even though they consist of integer-order components.

To overcome that challenge, the numerator and the denominator of $G_{\infty,\varnothing}(s)$ are regarded as two polynomials with respect to some functions of $s$ instead of the $s$ itself. For the infinite tree network, Eq.~\cref{eq:treeUndInf} reveals that the numerator and denominator of its $G_{\infty,\varnothing}(s)$ can be viewed as two polynomials where the variable is $\phi_1(s)=s^{\frac{1}{2}}$. While, for the infinite electrical ladder network, the numerator and the denominator of Eq.~\cref{eq:eLadderUndInf} are two bivariate polynomials whose variables are
\begin{align}
    \phi_1(s)&=s,\nonumber\\
    \phi_2(s)&=\sqrt{s^2+\cfrac{2r_1+4r_2}{r_1r_2c}s+\cfrac{r_1+4r_2}{r_1r_2^2c^2}}.\label{eq:eLadderInfPhi2}
\end{align}
Similarly, for the infinite mechanical ladder network, the numerator and the denominator of Eq.~\cref{eq:mLadderUndInf} are also two bivariate polynomials whose variables are
\begin{align*}
    \phi_1(s)&=s,\\
    \phi_2(s)&=A(s),
\end{align*}
where $A(s)$ is defined by Eq.~\cref{eq:mLadderUndInfA}.

Due to the assumptions (A-3) and (A-4), the recurrence formulas, $G_r(s)$, are always rational expressions whose numerator and denominator are polynomials in $s$. Hence, similar to finite networks, the numerator and denominator of an infinite network's transfer function $G_\infty(s)$, which is the result of repeatedly using $G_r(s)$ starting with $G_{\infty,\varnothing}(s)$, are also two polynomials whose variables are the union of the ones originating from those two sources, \textit{i.e.}
\begin{equation*}
    \{s\}\bigcup\{\phi_1(s),\phi_2(s),\dots\}.
\end{equation*}
Note that $\{s\}$ can be omitted from the above set if there exists an integer power of some polynomial variables of $G_{\infty,\varnothing}(s)$ equals $s$. For instance, the $\phi_1^2(s)=s$ for the tree network and $\phi_1^1(s)=s$ for both electrical and mechanical ladders. Therefore, the polynomial variables of $G_\infty(s)$ are in fact same as those of $G_{\infty,\varnothing}(s)$ for all three examples in this paper. However, suppose the polynomial variables for some infinite network's $G_{\infty,\varnothing}(s)$ are $\phi_1=\sqrt{s+1}$ and $\phi_2=\sqrt{s+2}$, where neither of them can equal $s$ for any integer powers. Then, we should include an additional polynomial variable $\phi_3(s)=s$ for $G_\infty(s)$.

The above results cause another difference between computing finite and infinite networks' transfer functions, which is the fact that the numerator and the denominator of $G_\infty(s)$ are no longer univariate. Consequently, we need to employ higher-dimensional convolutions. Note that the equivalence between $n$-variable polynomial multiplications and $n$-dimensional tensor convolutions holds for all positive integers $n$. However, because all three examples in this paper are at most bivariate, we only show that equivalence for matrix convolutions in this paper. For two bivariate polynomials
\begin{align*}
    a(\phi_1,\phi_2)&=a_{1,1}\phi_1^{n_a-1}\phi_2^0+\cdots+a_{1,n_a}\phi_1^0\phi_2^{n_a-1}+a_{2,2}\phi_1^{n_a-2}\phi_2^0+\cdots+a_{2,n_a}\phi_1^0\phi_2^{n_a-2}+\cdots\\
    &+a_{n_a,n_a}\phi_1^0\phi_2^0,\\
    b(\phi_1,\phi_2)&=b_{1,1}\phi_1^{n_b-1}\phi_2^0+\cdots+b_{1,n_b}\phi_1^0\phi_2^{n_b-1}+b_{2,2}\phi_1^{n_b-2}\phi_2^0+\cdots+b_{2,n_b}\phi_1^0\phi_2^{n_b-2}+\cdots\\
    &+b_{n_b,n_b}\phi_1^0\phi_2^0,
\end{align*}
we can define their coefficient matrices as
\begin{align}
    &c_a=\begin{bmatrix}a_{1,1}&a_{1,2}&\cdots&a_{1,n_a-1}&a_{1,n_a}\\0&a_{2,2}&\cdots&a_{2,n_a-1}&a_{2,n_a}\\0&0&\cdots&a_{3,n_a-1}&a_{3,n_a}\\\vdots&\vdots&\ddots&\vdots&\vdots\\0&0&\cdots&0&a_{n_a,n_a}\end{bmatrix},\nonumber\\
    &c_b=\begin{bmatrix}b_{1,1}&b_{1,2}&\cdots&b_{1,n_b-1}&b_{1,n_b}\\0&b_{2,2}&\cdots&b_{2,n_b-1}&b_{2,n_b}\\0&0&\cdots&b_{3,n_b-1}&b_{3,n_b}\\\vdots&\vdots&\ddots&\vdots&\vdots\\0&0&\cdots&0&b_{n_b,n_b}\end{bmatrix}.\label{eq:coeffMatDef}
\end{align}
Then, the matrix convolution result $c_c=c_a*c_b$ is the coefficient matrix for the multiplication result $c(\phi_1,\phi_2)=a(\phi_1,\phi_2)b(\phi_1,\phi_2)$. The definition of that matrix convolution is
\begin{equation*}
    c_c(j,k)=\sum_p\sum_qc_a(p,q)c_b(j-p+1,k-q+1),
\end{equation*}
where the indices $p$ and $q$ run over all values that lead to legal subscripts of $c_a(p,q)$ and $c_b(j-p+1,k-q+1)$.

Using vector and matrix convolutions, we are now able to rewrite $G_{\infty,\varnothing}(s)$ and recurrence formulas $G_r(s)$ for those three examples in a way similar to finite networks. For the infinite tree network, from its $G_{\infty,\varnothing}(s)$ in Eq.~\cref{eq:treeUndInf}, we know its coefficient vectors are
\begin{align}
    c_{N\infty,\varnothing}&=\begin{bmatrix}1\end{bmatrix},\label{eq:recTreeCoeffInfUnd1}\\
    c_{D\infty,\varnothing}&=\begin{bmatrix}\sqrt{kb}&0\end{bmatrix}.\label{eq:recTreeCoeffInfUnd2}
\end{align}
Additionally, its recurrence formula Eq.~\cref{eq:recTree} is converted to
\begin{align}
    c_{Nr}&=\begin{bmatrix}k_{1,1}b_{1,1}&0&0\end{bmatrix}*c_{Ns1}*c_{Ns2}+\begin{bmatrix}k_{1,1}\end{bmatrix}*c_{Ns1}*c_{Ds2}+\begin{bmatrix}b_{1,1}&0&0\end{bmatrix}*c_{Ns2}*c_{Ds1}\nonumber\\
    &+c_{Ds1}*c_{Ds2},\label{eq:recTreeCoeffInf1}\\
    c_{Dr}&=\begin{bmatrix}k_{1,1}b_{1,1}&0&0\end{bmatrix}*(c_{Ns1}*c_{Ds2}+c_{Ns2}*c_{Ds1})+\begin{bmatrix}b_{1,1}&0&k_{1,1}\end{bmatrix}*c_{Ds1}*c_{Ds2},\label{eq:recTreeCoeffInf2}
\end{align}
where $c_{Ns1}$, $c_{Ds1}$ are the coefficient vectors of the sub-network's $G_{s1}(s)$, and $c_{Ns2}$, $c_{Ds2}$ are those of $G_{s2}(s)$. Note that some slight differences between the above Eqs.~\cref{eq:recTreeCoeffInf1}, \cref{eq:recTreeCoeffInf2} and their counterparts for finite trees, Eqs.~\cref{eq:recTreeCoeff1} and \cref{eq:recTreeCoeff2}, are due to the fact that the polynomial variable here is $\phi_1(s)=s^{\frac{1}{2}}$ instead of the $s$ directly.

From the infinite electrical ladder network's $G_{\infty,\varnothing}(s)$ in Eq.~\cref{eq:eLadderUndInf}, following the definition of coefficient matrices~\cref{eq:coeffMatDef}, we can conclude that
\begin{align*}
    c_{N\infty,\varnothing}&=\begin{bmatrix}1&1\\0&\frac{1}{r_2c}\end{bmatrix},\\
    c_{D\infty,\varnothing}&=\begin{bmatrix}\frac{2}{r_1}&0\\0&\frac{2}{r_1r_2c}\end{bmatrix}.
\end{align*}
Likewise, its recurrence formula Eq.~\cref{eq:recELadder} is transformed into a format supporting coefficient computations where
\begin{align*}
    c_{Nr}&=\begin{bmatrix}r_{1,1}r_{1,2}c_1&0\\0&r_{1,1}+r_{1,2}\end{bmatrix}*c_{Ns1}+r_{1,1}r_{1,2}c_{Ds1},\\
    c_{Dr}&=\begin{bmatrix}r_{1,2}c_1&0\\0&1\end{bmatrix}*c_{Ns1}+r_{1,2}c_{Ds1}.
\end{align*}
Analogous to the plus signs in \Cref{sec:finModAna}, the additions here can also be performed between two matrices with different sizes, whose definition is consistent with the addition between two bivariate polynomials. For example,
\begin{equation*}
    \begin{bmatrix}1&2\\0&3\end{bmatrix}+\begin{bmatrix}4&5&6\\0&7&8\\0&0&9\end{bmatrix}=\begin{bmatrix}4&5&6\\0&8&10\\0&0&12\end{bmatrix}.
\end{equation*}

For the infinite mechanical ladder's $G_{\infty,\varnothing}(s)$ in Eq.~\cref{eq:mLadderUndInf}, its coefficient matrices are
\begin{align*}
    c_{N\infty,\varnothing}&=\begin{bmatrix}-m&0&0\\0&-b&1\\0&0&0\end{bmatrix},\\
    c_{D\infty,\varnothing}&=\begin{bmatrix}2mk_d&0&0&0\\0&2(mk_p+bk_d)&0&0\\0&0&2(mk_i+bk_p)&0\\0&0&0&2bk_i\end{bmatrix}.
\end{align*}
Moreover, its recurrence formula~\cref{eq:recMLadder} can be re-written as
\begin{align*}
    c_{Nr}&=\begin{bmatrix}k_{d1}&0&0\\0&k_{p1}&0\\0&0&k_{i1}\end{bmatrix}*c_{Ns1}+\begin{bmatrix}1&0\\0&0\end{bmatrix}*c_{Ds1},\\
    c_{Dr}&=\begin{bmatrix}m_1&0&0\\0&b_1&0\\0&0&0\end{bmatrix}*\begin{bmatrix}k_{d1}&0&0\\0&k_{p1}&0\\0&0&k_{i1}\end{bmatrix}*c_{Ns1}+\begin{bmatrix}m_1&0&0&0\\0&b_1+k_{d1}&0&0\\0&0&k_{p1}&0\\0&0&0&k_{i1}\end{bmatrix}*c_{Ds1}.
\end{align*}

Similar to the modeling procedure for finite networks' transfer functions, after converting $G_{\infty,\varnothing}(s)$ and $G_r(s)$ into a form supporting pure coefficient computations, we use a recursive algorithm listed in \Cref{alg:anaInf} to obtain an infinite network's transfer function. The structure of that recursive algorithm is same as \Cref{alg:numInf}, while its feature of coefficient computations is same as \Cref{alg:anaFin}. Especially, the \texttt{CUnd()} function returns the coefficient tensors for $G_{\infty,\varnothing}(s)$, such as Eqs.~\cref{eq:recTreeCoeffInfUnd1} and \cref{eq:recTreeCoeffInfUnd2} for infinite trees, and the \texttt{Cr()} function contains the convolution computations for recurrence formulas, \textit{i.e.}, Eqs.~\cref{eq:recTreeCoeffInf1} and \cref{eq:recTreeCoeffInf2} for infinite trees.

\begin{algorithm}
    \caption{Pseudocode of our modeling algorithm for infinite networks' transfer functions. It computes the coefficient tensors \texttt{cN} and \texttt{cD} of an infinite network's transfer function given its damage case \texttt{(l,e)} and the undamaged constants \texttt{undCst}.}
    \label{alg:anaInf}
    \begin{algorithmic}[1]
    \STATE{\texttt{\textbf{function}~[cN,cD]~=~tranInf(l,e,undCst)}}
    \IF{\texttt{isEmpty(l)}}
    \STATE{\texttt{[cN,cD]~=~CUnd(undCst);}}
    \ELSE
    \STATE{\texttt{[l1,e1,lS,eS]~=~partition(l,e);}}
    \FOR{\texttt{idx from 1 to nS}}
    \STATE{\texttt{[cNS[idx],cDS[idx]]~=~tranInf(lS[idx],eS[idx],undCst);}}
    \ENDFOR
    \STATE{\texttt{g1Cst~=~getG1Cst(l1,e1,undCst);}}
    \STATE{\texttt{[cN,cD]~=~Cr(g1Cst,cNS,cDS);}}
    \STATE{\texttt{[cN,cD]~=~simplify(cN,cD);}}
    \ENDIF
    \end{algorithmic}
\end{algorithm}

By using \Cref{alg:anaInf}, we can obtain that for an infinite tree, its transfer function is
\begin{equation*}
    G_{\infty,(\boldsymbol{l},\boldsymbol{\epsilon})}(s)=\frac{0.71s^2+2.20s^{\frac{3}{2}}+9.62s+12.20s^{\frac{1}{2}}+1.41}{s^{\frac{5}{2}}+3.11s^2+13.60s^{\frac{3}{2}}+3.40s+2.00s^{\frac{1}{2}}},
\end{equation*}
when the damage case is $(\boldsymbol{l},\boldsymbol{\epsilon})=([k_{2,1},b_{2,1}],[0.1,0.2])$. For an infinite electrical ladder whose damage case is $(\boldsymbol{l},\boldsymbol{\epsilon})=([r_{2,2},r_{3,2}],[0.1,0.1])$, the analytical expression of its input impedance is
\begin{equation*}
    G_{\infty,(\boldsymbol{l},\boldsymbol{\epsilon})}(s)=\frac{N_{\infty,(\boldsymbol{l},\boldsymbol{\epsilon})}(s)}{D_{\infty,(\boldsymbol{l},\boldsymbol{\epsilon})}(s)},
\end{equation*}
where
\begin{align*}
    N_{\infty,(\boldsymbol{l},\boldsymbol{\epsilon})}(s)&=s^4+s^3\phi_2(s)+7.2\times10^3s^3+5.2\times10^3s^2\phi_2(s)+1.5\times10^7s^2+6.7\times10^{6}s\phi_2(s)\\
    &+8.1\times10^9s+1.5\times10^9\phi_2(s)+8.0\times10^{10},\\
    D_{\infty,(\boldsymbol{l},\boldsymbol{\epsilon})}(s)&=0.1s^4+0.1s^3\phi_2(s)+622s^3+421s^2\phi_2(s)+9.8\times10^5s^2+3.5\times10^5s\phi_2(s)\\
    &+2.6\times10^8s+2.2\times10^7\phi_2(s)+2.5\times10^9,
\end{align*}
and $\phi_2(s)$ is defined by Eq.~\cref{eq:eLadderInfPhi2}. For an infinite mechanical ladder network whose damage case is $(\boldsymbol{l},\boldsymbol{\epsilon})=([k_{p2},k_{i2},k_{d2}],[0.1,0.1,0.1])$, the coefficient matrices for its transfer function are
\begin{align*}
    &c_{N\infty,(\boldsymbol{l},\boldsymbol{\epsilon})}=\begin{bmatrix}1&0&0&0&0&0&0&0\\0&9.21&0.05&0&0&0&0&0\\0&0&35.6&0.42&0&0&0&0\\0&0&0&84.0&1.33&0&0&0\\0&0&0&0&62.1&2.70&0&0\\0&0&0&0&0&5.66&0.26&0\\0&0&0&0&0&0&0.14&0.01\\0&0&0&0&0&0&0&0\end{bmatrix},\\
    &c_{D\infty,(\boldsymbol{l},\boldsymbol{\epsilon})}=\begin{bmatrix}1&0&0&0&0&0&0&0&0&0\\0&12.2&0.05&0&0&0&0&0&0&0\\0&0&69.2&0.58&0&0&0&0&0&0\\0&0&0&219&2.91&0&0&0&0&0\\0&0&0&0&311&7.76&0&0&0&0\\0&0&0&0&0&217&5.91&0&0&0\\0&0&0&0&0&0&74.8&0.54&0&0\\0&0&0&0&0&0&0&8.63&0.01&0\\0&0&0&0&0&0&0&0&0.40&0\\0&0&0&0&0&0&0&0&0&0.01\end{bmatrix}.
\end{align*}

\section{Discussion}
\label{sec:dis}
In this section, we discuss some discoveries, applications and thoughts regarding the modeling results obtained by the procedures in \Cref{sec:finMod,sec:infMod}. In \Cref{sec:disMul}, we isolate the effects brought by changing a network's condition on its dynamics. This isolation is important as it suggests that those networks can be further studied using robust control tools when the variations in the conditions are uncertain. \Cref{sec:disFrac} explains the reason why infinite networks always behave in a non-integer-order manner from the perspective of the zeros and poles in finite networks' transfer functions. \Cref{sec:disRal} illustrates that by using finite networks' transfer functions, we can approximate some irrational functions via rational expressions. Finally, \Cref{sec:disFin} discusses the rationale behind choosing whether to model a finite network's response accurately or to approximate it with the corresponding infinite one's.

\subsection{Effects of varying a network's status on its dynamics}
\label{sec:disMul}
One crucial discovery after writing down a network's transfer function is that the effect of varying its status on its dynamics can be isolated. That isolation is described as a multiplicative disturbance in this paper, which is one of the classical models in the robust control research area.

Specifically, from previous sections, we confirm that a self-similar network's transfer function is always a ratio between two functions of $s$. Therefore, if at two different statuses $a$ and $b$, a network's respective transfer functions are
\begin{align*}
    G_a(s)&=\frac{N_a(s)}{D_a(s)},\\
    G_b(s)&=\frac{N_b(s)}{D_b(s)},
\end{align*}
we then can express the effect of changing that network's status from $a$ to $b$ in regard to its transfer function as a multiplicative disturbance $\Delta(s)$ where
\begin{equation*}
    \Delta(s)=\frac{G_b(s)}{G_a(s)}=\frac{N_b(s)D_a(s)}{N_a(s)D_b(s)}.
\end{equation*}

There exist at least two meaningful perspectives that can be explored in the context of large networks. The first one is quantifying the approximation error of estimating a finite network's transfer function using the corresponding infinite one's. As an example, for an undamaged tree network, we know that
\begin{align*}
    G_{\infty,\varnothing}(s)&=\frac{1}{1.4142\sqrt{s}},\\
    G_{3,\varnothing}(s)&=\frac{3s^6+62s^5+428s^4+1272s^3+1712s^2+992s+192}{s^7+30s^6+300s^5+1288s^4+2576s^3+2400s^2+960s+128}.
\end{align*}
Then, the error of approximating a three-generation tree's transfer function using the infinite tree's is
\begin{equation*}
    G_{3,\varnothing}(s)=G_{\infty,\varnothing}(s)\frac{4.243s^{\frac{13}{2}}+87.68s^{\frac{11}{2}}+605.3s^{\frac{9}{2}}+1799s^{\frac{7}{2}}+2421s^{\frac{5}{2}}+1403s^{\frac{3}{2}}+271.5s^{\frac{1}{2}}}{s^7+30s^6+300s^5+1288s^4+2576s^3+2400s^2+960s+128}.
\end{equation*}
The second perspective is to study the effect brought by a network's damages on its response. Again, for an infinite tree whose damage case is $(\boldsymbol{l},\boldsymbol{\epsilon})=([k_{2,1},b_{2,1}],[0.1,0.2])$, its frequency response is
\begin{equation*}
    G_{\infty,(\boldsymbol{l},\boldsymbol{\epsilon})}(s)=\frac{0.7071s^2+2.2s^{\frac{3}{2}}+9.6167s+12.2s^{\frac{1}{2}}+1.4142}{s^{\frac{5}{2}}+3.1113s^2+13.6s^{\frac{3}{2}}+3.3941s+2s^{\frac{1}{2}}}.
\end{equation*}
As a result, the effect brought by that damage case is
\begin{equation*}
    G_{\infty,(\boldsymbol{l},\boldsymbol{\epsilon})}(s)=G_{\infty,\varnothing}(s)\frac{s^2+3.1113s^{\frac{3}{2}}+13.6s+17.2534s^{\frac{1}{2}}+2}{s^2+3.1113s^{\frac{3}{2}}+13.6s+3.3941s^{\frac{1}{2}}+2}.
\end{equation*}

\subsection{Fractional or irrational nature of infinite networks' dynamics}
\label{sec:disFrac}
Knowing transfer functions for finite networks helps explaining the non-integer-order nature of infinite networks' dynamics from the perspective of zeros and poles. That non-integer-order behavior is consistent with other infinite dimensional systems as suggested by literature. \cite{TRIGEASSOU2013892}

\begin{figure}
    \centering
    \includegraphics[width=.75\textwidth]{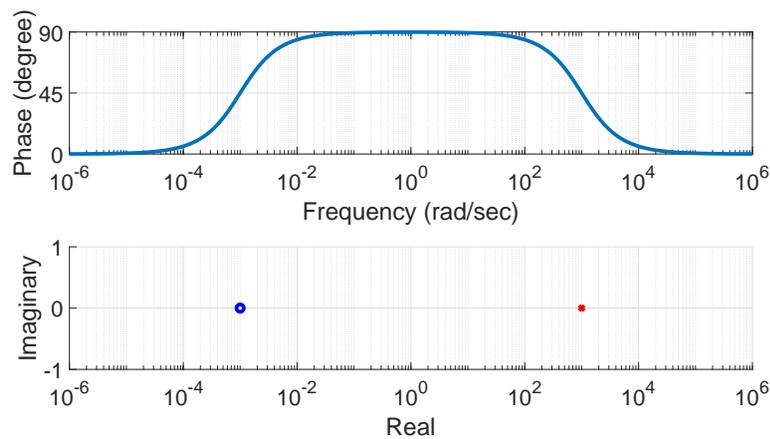}
    \caption{The effect brought by a zero and a pole on the phase of a dynamical system's frequency response. The transfer function used in this figure is $G(s)=\frac{s+10^{-3}}{s+10^3}$.}
    \label{fig:effectOfZP1}
\end{figure}

\begin{figure}
    \centering
    \includegraphics[width=.75\textwidth]{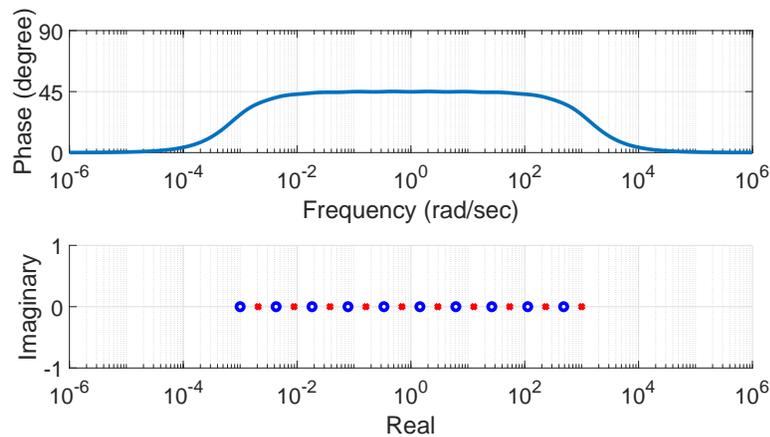}
    \caption{An integer-order system behaves like a fractional-order system within the region where lots of zeros and poles concentrate. The example dynamical system in this figure would behave like a half-order on within the frequency bandwidth where the phase is $45^o$.}
    \label{fig:effectOfZP2}
\end{figure}

The effect brought by a zero and a pole on a frequency response's phase is shown in \Cref{fig:effectOfZP1}, where a zero increases the phase by $90^o$ while a pole decreases it by $90^o$, and there exists a transitional region where that zero and pole locates. As a result, usual integer-order systems' phase always stays at multiples of $90^o$ for a wide bandwidth of frequencies. It is not difficult to understand that when lots of zeros and poles concentrate within one region, the dynamical system should behave in a non-integer-order manner in that bandwidth of frequency, as illustrated in \Cref{fig:effectOfZP2}.

\begin{figure}
    \centering
    \includegraphics[width=.75\textwidth]{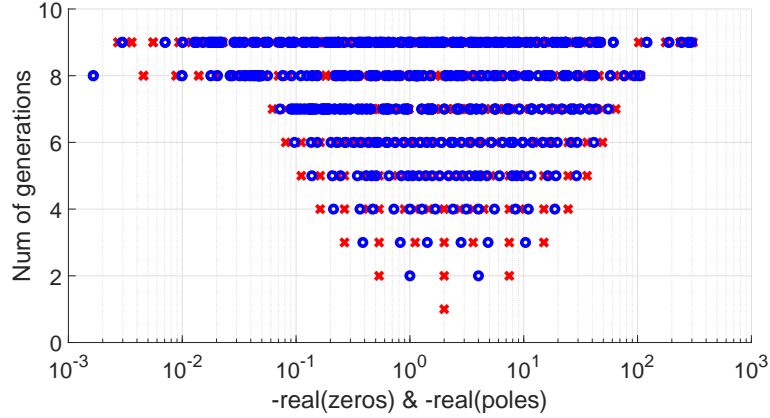}
    \caption{Negative of the zeros' and poles' real parts for finite undamaged trees from one-generation to nine-generation. The blue circles are for zeros and the red crosses are for poles. Note that those zeros and poles with positive real parts are ignored in this figure since the $x$-axis is plotted in the $log$-scale.}
    \label{fig:zPFinTree}
\end{figure}

That concentration of zeros and poles for an integer-order system renders its behavior like a non-integer-order one is exactly what happens to a finite network's dynamics as it grows larger. As a concrete example, using \Cref{alg:anaFin}, we obtain the transfer functions of undamaged finite trees from the one-generation network to the nine-generation one, that is from $G_{1,\varnothing}(s)$ to $G_{9,\varnothing}(s)$. \Cref{fig:zPFinTree} plots the negative of their zeros' and poles' real parts as the number of generations increases, from which we can observe the above-mentioned effect brought by piling up zeros and poles. Note that the region where zeros and poles pile up is consistent with the transitional region in \Cref{fig:treeUndInf} where phase is around $-45^o$. Hence, although finite networks' dynamics are always integer-order, as the number of their generations increases, their behaviors would eventually converge to non-integer-order dynamics for the corresponding infinite networks.

\subsection{Rational approximation of irrational functions}
\label{sec:disRal}
\begin{figure}
    \centering
    \includegraphics[width=.75\textwidth]{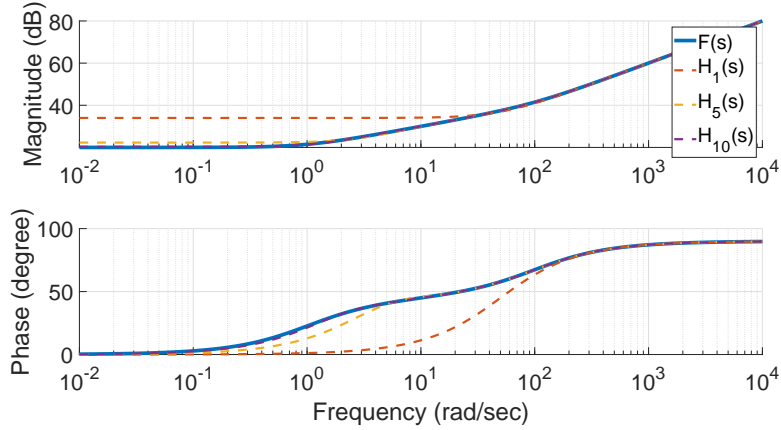}
    \caption{Use rational expressions $H_g(s)$ in Eq.~\cref{eq:approxIrH} to approximate the irrational function $F(s)$ in Eq.~\cref{eq:approxIrF} when $g=1,5,10$.}
    \label{fig:approxIr}
\end{figure}

Because finite networks' transfer functions are rational expressions while the limiting case of an infinite network contains irrational terms, it is straightforward to leverage that for a rational approximation of some irrational functions. Similar explorations can be found in literature such as \cite{wing1998rational}, from which we pick the following example: We want to approximate the following irrational functions
\begin{equation}
    F(s)=\sqrt{s^2+100s+100},
    \label{eq:approxIrF}
\end{equation}
which is a part of the undamaged transfer function for infinite electrical ladder, $G_{\infty,\varnothing}(s)$, in Eq.~\cref{eq:eLadderUndInf}. Since we know that $G_{g,\varnothing}(s)$ converges to $G_{\infty,\varnothing}(s)$ as $g$ goes to infinity, we can then use the rational function
\begin{equation}
    H_g(s)=G_{g,\varnothing}(s)\left(\frac{2}{r_1}s+\frac{2}{r_1r_2c}\right)-s-\frac{1}{r_2c},
    \label{eq:approxIrH}
\end{equation}
to approximate that irrational function $F(s)$. Comparing Eq.~\cref{eq:approxIrF} to its counterpart in Eq.~\cref{eq:eLadderUndInf}, we see that we need to find the undamaged constants satisfying the following two equations,
\begin{align*}
    &\frac{2r_1+4r_2}{r_1r_2c}=100,\\
    &\frac{r_1+4r_2}{r_1r_2^2c^2}=100.
\end{align*}
We can choose that $r_1=1$, $r_2=5\sqrt{6}+12$, and $c=\frac{5+2\sqrt{6}}{120+50\sqrt{6}}$, and plug them into the algorithm to obtain the coefficients for $G_{g,\varnothing}(s)$, which leads to the analytical expression for the rational function $H_g(s)$ according to Eq.~\cref{eq:approxIrH}. For instance, we can choose the number of generation $g=5$ which leads to
\begin{equation*}
    G_{5,\varnothing}(s)=\frac{s^5+226s^4+1.8\times10^4s^3+5.7\times10^5s^2+6.5\times10^6s+1.5\times10^7}{s^5+201s^4+1.3\times10^4s^3+3.3\times10^5s^2+2.4\times10^6s+2.1\times10^6},
\end{equation*}
That would give us the following $H_5(s)$ as one rational approximation for $F(s)$,
\begin{equation*}
    H_5(s)=\frac{s^6+251s^5+2.2\times10^4s^4+8.2\times10^5s^3+1.1\times10^7s^2+3.8\times10^7s+2.8\times10^7}{s^5+201s^4+1.3\times10^4s^3+3.3\times10^5s^2+2.4\times10^6s+2.1\times10^6}.
\end{equation*}
\Cref{fig:approxIr} shows the rational expression $H_g(s)$ when $g=1,5,10$, from which we can conclude that $H_{10}(s)$ is already a reasonable approximation of the irrational function $F(s)$ in Eq.~\cref{eq:approxIrF}. Note that the trend of convergence in \Cref{fig:approxIr} agrees with the discussion in \Cref{sec:disFrac}. When there is only one generation, $H_1(s)$ is already same as $F(s)$ within the integer-order behavior region where the phase is $0^o$ or $90^o$. However, in the transitional region between those two phases, it requires multi-generation networks to obtain better approximation results as the zeros and poles gradually pile up within that region.

\subsection{Modeling finite networks exactly versus approximation using infinite networks}
\label{sec:disFin}
When choosing between whether to model a finite network's frequency response directly or to approximate it with its corresponding infinite version, there usually exists a trade-off between two considerations: accuracy and computation time. It is very clear that the finite modeling algorithm always returns an accurate result for a finite network's response, but it requires more computational time as that network grows larger. Comparing the modeling algorithms for finite networks to those for infinite networks, we can see that the running time of finite networks' modeling algorithm is mainly determined by the number of generations inside that network, while that of infinite networks' is decided by the deepest generation where damages reside. Therefore, when we have a very large finite network which is undamaged or whose damages only happen at shallow generations, approximating its frequency response using the corresponding infinite network's may be a good alternative.

Take a $20$-generation tree as an example where damages only happen in the first three generations: $(\boldsymbol{l},\boldsymbol{\epsilon})=([k_{2,1},k_{2,2},b_{3,1}],[0.1,0.2,0.3])$. From \Cref{fig:treeNumInf}, we see that its frequency response $G_{20,(\boldsymbol{l},\boldsymbol{\epsilon})}(s)$ overlaps the corresponding infinite network's $G_{\infty,(\boldsymbol{l},\boldsymbol{\epsilon})}(s)$ in $3.8$ decades of frequencies from $0.02rad/sec$ to $120rad/sec$. However, the computation time of $G_{20,(\boldsymbol{l},\boldsymbol{\epsilon})}(s)$ is $16$ seconds, while $G_{\infty,(\boldsymbol{l},\boldsymbol{\epsilon})}(s)$ only takes $0.01$ seconds. Incidentally, because infinite trees' frequency responses display non-integer-order behavior across the entire frequency range, its approximation ability of finite ones is relatively weak. In contrast, if an infinite network only behaves in a non-integer fashion for some frequencies, its ability of approximating finite networks is much stronger, which can be seen in \Cref{fig:eLadderNumInf,fig:mLadderNumInf} for electrical and mechanical ladders. For example, in \Cref{fig:eLadderNumInf}, the finite electrical ladder's response $G_{20,(\boldsymbol{l},\boldsymbol{\epsilon})}(s)$ overlaps the infinite one's $G_{\infty,(\boldsymbol{l},\boldsymbol{\epsilon})}(s)$ for almost all frequencies. However, $G_{20,(\boldsymbol{l},\boldsymbol{\epsilon})}(s)$ takes $0.2$ seconds to compute and $G_{\infty,(\boldsymbol{l},\boldsymbol{\epsilon})}(s)$ only needs $0.009$ seconds.

\section{Concluding remarks}
\label{sec:con}
In this work, we focus on computing frequency response and transfer functions for large self-similar networks, which are classified in two ways: First, a network is either finite or infinite. Second, a network is either damaged or intact. For all combinations of different conditions, we illustrate the algorithms and the procedures to evaluate both the corresponding frequency response and transfer functions, which are applied to three example networks illustrating their capabilities of handling a general class of large self-similar networks. Building upon that main result, the following three points are also highlighted in this paper. First, the effect of varying a network's condition on its response can be isolated complying with a formation commonly used in the robust control research area. Second, we explain the non-integer-order nature observed in infinite dimensional systems' dynamics from the perspective of the zeros and poles of finite networks' transfer functions. Third, we also leverage the analytical expressions of finite networks' rational transfer functions to approximate some irrational expressions, which concerns the realization of non-integer-order dynamics.


\bibliographystyle{siamplain}
\bibliography{references}

\end{document}